\documentclass[showpacs, preprintnumbers, amsmath, amssymb, superscriptaddress, prb, onecolumn, aps]{revtex4-2}
\usepackage{color}
\usepackage{amsmath, amssymb, amsfonts, mathtools}
\usepackage{graphicx}
\usepackage{float}
\usepackage{bm}
\usepackage{multirow}
\usepackage[caption=false]{subfig}
\usepackage[colorlinks]{hyperref}
\usepackage{physics}
\usepackage{bbold}
\usepackage{pifont}
\usepackage{tikz}
\usepackage{dcolumn}
\usepackage{placeins}

\newcommand{\vect}[1]{\mathbf{#1}}

\DeclareMathAlphabet{\bi}{OML}{cmm}{b}{it}

\def\be{\begin{equation}}
\def\ee{\end{equation}}
\def\bearr{\begin{eqnarray}}
\def\eearr{\end{eqnarray}}

\begin{document}
	
	\title{Topological Properties of Bilayer $\alpha$-T$_3$ Lattice Induced by Polarized Light}
	
	\author{O.~Benhaida}
	\affiliation{LPHE, Modeling and Simulations, Faculty of Science, Mohammed V University in Rabat, Rabat, Morocco}
	\affiliation{CPM, Centre of Physics and Mathematics, Faculty of Science, Mohammed V University in Rabat, Rabat, Morocco}
	
	\author{E.~H.~Saidi}
	\affiliation{LPHE, Modeling and Simulations, Faculty of Science, Mohammed V University in Rabat, Rabat, Morocco}
	\affiliation{CPM, Centre of Physics and Mathematics, Faculty of Science, Mohammed V University in Rabat, Rabat, Morocco}
	\affiliation{College of Physical and Chemical Sciences, Hassan II Academy of Sciences and Technology, Rabat, Morocco}
	
	\author{L.~B.~Drissi}
	\email[Corresponding author: ]{ldrissi@fsr.ac.ma} 
	\affiliation{LPHE, Modeling and Simulations, Faculty of Science, Mohammed V University in Rabat, Rabat, Morocco}
	\affiliation{CPM, Centre of Physics and Mathematics, Faculty of Science, Mohammed V University in Rabat, Rabat, Morocco}
	\affiliation{College of Physical and Chemical Sciences, Hassan II Academy of Sciences and Technology, Rabat, Morocco}
	
	\author{R.~Ahl~Laamara}
	\affiliation{LPHE, Modeling and Simulations, Faculty of Science, Mohammed V University in Rabat, Rabat, Morocco}
	\affiliation{CPM, Centre of Physics and Mathematics, Faculty of Science, Mohammed V University in Rabat, Rabat, Morocco}	
\begin{abstract}
\vspace{0.1cm}
We investigate the topological properties of photon-dressed energy bands in bilayer $\alpha-T_{3}$ lattices under off-resonant circularly polarized light, focusing on aligned and cyclic stacking configurations. Analytical expressions for quasi-energy bands are derived for aligned stacking, while numerical results address cyclic stacking at Dirac points. Circularly polarized light breaks the time-reversal symmetry, lifting the degeneracies at the intersections $t^{a,c}$, leading to the appearance of a Haldane-type Chern insulator in the absence of a magnetic field . At $\alpha = 1/\sqrt{2}$, orbital magnetic moments of corrugated and flat bands exhibit opposite signs, as do their Berry curvatures. For $0 < \alpha < 1$, light-induced band deformations near Dirac points create gaps in the quasi-energy spectrum, where the chemical potential modulates orbital magnetization. Linear magnetization variations align with Chern numbers, yielding quantized anomalous Hall conductivity across stacking types. Notable particle-hole symmetry breaking within $0 < \alpha < 1$ suggests applications in valley caloritronics and quantum sensing. At $\alpha = 1$, flat and corrugated bands remain undistorted; while the flat band contributes no Berry curvature, it produces a finite negative orbital magnetic moment, contrasting with the positive moment of the corrugated band.
\end{abstract}
\maketitle
\section{ Introduction}

Topological phases of matter have garnered significant interest revealing a
rich tapestry of phenomena that challenge our understanding of condensed
matter systems \cite{S1 16}-\cite{114}. The interplay between topology and
symmetry leads to exotic states , such as quantum Hall insulators and
topological insulators, where edge states are robust against disorder \cite%
{115,116,117}. These edge states are protected by the topological invariants
that characterize the bulk properties of the material, making them resilient
to perturbations that would typically localize or scatter electronic states
\cite{118,119}. In addition to these well-known phases, Weyl semimetals have
emerged as a new class of materials exhibiting unique features such as Fermi
arcs on their surfaces \cite{120}. Higher-order topological insulators
further enrich this landscape by exhibiting boundary modes localized at
corners or hinges, providing a new dimension to the study of topological
phenomena \cite{121,122,123}. Furthermore, the discovery of Majorana bound
states in topological superconductors has opened pathways for fault-tolerant
quantum computing, as these states can encode quantum information in a
non-local manner \cite{124,125}.

Periodically driven electronic and photonic systems have seen remarkable
advancements in recent years, leading to the discovery of new phenomena and
applications across various fields. The non-equilibrium dynamics underlying
these systems, as described by Floquet theory \cite{30in,31in}, have emerged
as a powerful tool for investigating non-trivial topological phases in
two-dimensional materials \cite{In2Da,41in,42in,44in,in2Db}. Applying
Floquet engineering to photonic crystals allows the observation of the
anomalous quantum Hall effect by imaging topological edge states that arise
from periodic driving \cite{20in}. Thermalization processes under periodic
driving lead to unique dynamics that differ from equilibrium systems \cite%
{22in}. Unconventional superconductivity in twisted bilayer graphene (TBG)
at a magic angle \cite{32in} and in TBG aligned with boron nitride,
demonstrating isolated flat bands under external electric fields \cite%
{35in,36in} complemented this landscape with respect to Conventional 2D
superconductors \cite{ilyas}. These findings not only enhance our
understanding of topological phases but also pave the way for innovative
applications in quantum technologies and materials science.

\textbf{Interestingly,} the light-induced Hall effect in graphene has
stimulated extensive research into Floquet topological phases \cite{38in}.
This area of study encompasses a variety of phenomena, including topological
phase transitions and the emergence of Floquet spin states \cite{46in}.
These investigations have revealed a range of exotic behaviors that
challenge conventional understandings of quantum materials\textbf{\ }\cite%
{49in,50in}. A crucial mechanism underlying these effects is the
modification of the band structure through virtual photon processes, which
occur under off-resonant circularly polarized light irradiation. This is
also exemplified by the emergence of flat bands in systems such as Kagome
\cite{56in}, $T_{3}$ or Dice \cite{58in}, and Lieb lattices \cite{60in}. Of
particular note is the fact that the $T_{3}$ or Dice lattice exhibits flat
bands in the vicinity of charge neutrality, which results in electronic
bands devoid of dispersion \cite{33in,61in}. The dice lattice structure can
be realized through various experimental techniques. Traditionally, it has
been constructed by epitaxially growing three layers of cubic lattices, such
as $SrTiO_{3}/SrIrO3/SrTiO_{3}$, along the (111) crystallographic direction
\cite{63in}. \newline
Alternatively, optical lattice techniques have been proposed, where the
interference of three pairs of counter-propagating laser beams at a
wavelength of $\lambda =3a/2$ (a being the lattice constant) can generate
the dice lattice geometry. Theoretical proposals have been made to realize the $\alpha-T_{3}$ optical lattice, for example, by introducing a phase difference between one of the laser pairs \cite{64in,65in}. This approach
offers a degree of tunability as the parameter $\alpha $, which controls the
lattice geometry, can be continuously modified by adjusting the laser phase.
Furthermore, the low-energy behavior of certain quantum well systems, such
as $Hg_{1}-xCdxTe$, can be effectively mapped onto the $\alpha -T_{3}$ model
for appropriate doping levels, with $\alpha $ taking the specific value of $%
1/\sqrt{3}$ \cite{a66in}. The $\alpha -T_{3}$ lattice has attracted
significant research interest due to its tunable band structure and
potential for realizing novel topological states \cite{69in,a69in,70in}. A key
feature of the $\alpha -T_{3}$ model is the continuously varying Berry phase
from $\pi $ (graphene) to 0 (dice lattice) as the parameter $a\alpha $ is
adjusted \cite{71in}. This unique characteristic allows for the study of how
the Berry phase influences various physical phenomena. Several studies have
explored the dependence of transport and optical properties on the Berry
phase in the $\alpha -T_{3}$ lattice. The impact of $\alpha $ on the quantum
Hall conductivity, dynamic longitudinal photoconductivity, and SdH
oscillations was demonstrated in \cite{71in}. Additionally, the effect of
varying Berry phase on orbital magnetostriction has been observed \cite{65in}%
. Transport measurements by Biswas and al. \cite{69in} revealed a smooth
transition in the Hall conductivity from $\sigma _{xy}=2(2n+1)(e^{2}/h)$ to $%
\sigma _{xy}=4n(e^{2}/h)$ with increasing $\alpha $ (n being an integer),
highlighting the interplay between the model parameter and the topological
behavior. Studies have explored the interplay between Floquet states and the
Berry phase in the $\alpha -T_{3}$ lattice irradiated with circularly
polarized resonant light \cite{74in}. This work highlighted the crucial role
of resonance in inducing a Berry phase-dependent optical gap. Conversely,
off-resonant circularly polarized radiation has been demonstrated to open
gaps in various material systems, including graphene \cite{75in},
topological insulator surface states \cite{75in}, silicene \cite{76in},
quasi-Dirac systems \cite{77in} and MoS2 \cite{78in}. This process
transforms these materials into Chern insulating states. It is important to
note that, as emphasized by Kitagawa et al. \cite{42in}, the multiple
Floquet bands formed by resonant light cannot be directly treated as a new,
static band structure for calculating transport properties in these
non-equilibrium systems. Notably, even higher Chern numbers were observed in
single-layer Dice lattices \cite{81in}. A Haldane-like model on a Dice
lattice predicted Chern numbers of $\pm 2$ for the dispersive valence and
conduction bands and 0 for the flat middle band. Additionally, this system
exhibited a quantum anomalous Hall effect with two chiral edge channels \cite%
{81in}. The $\alpha -T_{3}$ lattice exhibits a unique behavior near charge
neutrality. While the flat bands become dispersed when touching
high-symmetry points (K and K$^{^{\prime }}$ in the presence of spin-orbit
coupling, they retain their non-trivial character \cite{83in}. Recent work
by Tamang et al. \cite{84in} explored the topological properties of the $%
\alpha -T_{3}$ lattice using off-resonant circularly polarized light and
found that the Berry curvature and orbital magnetic moment associated with
the flat band change signs at $\alpha =1/\sqrt{2}$. Moreover, the slopes of
the linear regions of the orbital magnetization were shown to be closely
related to the Chern numbers on either side of this critical $\alpha $
value. When two layers of the $\alpha -T_{3}$ lattice are stacked on top of
each other to form a bilayer lattice with unequal stacking, the low-energy
effective model exhibits a dispersive nature of the flat bands near the
charge neutrality point \cite{85in}. The effect of Haldane flow on this
lattice has also been analyzed, as it separates the six bands and determines
non-zero Chern numbers for each band, giving the system distinctive
topological properties. Furthermore, continuous modification of the scaling
parameter $\alpha $ leads to topological phase transitions that occur
through the band crossings, causing discontinuous changes in the Chern
numbers of the lower conduction and valence bands, where these changes
depend on the next-nearest-neighbor jump strength and appear at specific
values of the scaling parameter $\alpha $\cite{86in}. To date, no studies
have investigated the topological signatures of bilayer $\alpha -T_{3}$ lattices under irradiation.
\par This study investigates emerging topological features in the $\alpha-T_{3}$ bilayer for aligned and cyclic stacks subjected to circularly polarized light in the off-resonance regime. This irradiation induces an effective mass term in the Hamiltonian, which breaks the time-reversal symmetry and transforms the semi-metallic behaviour of these stacks into a Haldane-type Chern insulator near Dirac points.To analyze these topological transitions, we propose to calculate Berry curvature, orbital magnetic moment, orbital magnetization, and anomalous Hall conductivity. These physical quantities are sensitive indicators of the system's topology. For example, the Berry curvature associated with corrugated and flat bands changes sign when the control parameter $\alpha$ is varied around the critical value $1/\sqrt{2}$. The orbital magnetic moment shows a similar inversion at this same value of $\alpha$, clearly indicating the existence of a topological phase transition. In the presence of gaps, the orbital magnetization exhibits linear behaviour, which is a topological signature. These linear regions correlate with Chern numbers. Furthermore, the anormal Hall conductivity also adopts a topological behaviour when the chemical potential within the gaps varies, leading to the appearance of plateaus. These plateaus appear when the parameters $\alpha$ and light amplitude (the latter being proportional to the mass term) are controlled and are also observed in the Haldane model applied to cyclic stacking, as proposed by  P.Parui\cite{86in}. By modulating the electron-photon coupling, it is possible to realize the Haldane model in an $\alpha-T_{3}$ bilayer. The broken particle-hole symmetry, observed through properties, leads to valley anomalies for $0<\alpha<1$. In addition, the behaviour of the polarized valley current has also been studied under the effect of circularly polarized light irradiating a monolayer $\alpha-T_{3}$ lattice, with the aim of investigating the topological properties of this current \cite{87in,84in}. These results are promising and could pave the way for future applications in valley caloritronics and quantum sensing.
\newline

This paper is organized as follows: In Section \ref{sec1}, we present a
Hamiltonian model for the bilayer system in both aligned and cyclic stacks
and outline the method for extracting the effective Hamiltonian induced by
circularly polarized light, focusing on time-reversal and particle-hole
symmetries. In Section \ref{sec1}, the results from Section \ref{subsec1}
are discussed in the context of quasi-energy. Sections \ref{subsec2} , \ref%
{subsec3} and \ref{subsec4} explore topological features, including Berry
curvature, the orbital magnetic moment, and orbital magnetization.
Additionally, Section \ref{subsec5} provides an analysis of the anomalous
Hall conductivity. Finally, the paper concludes with a comprehensive summary
of the results presented in Section \ref{sec2}.

\section{ THEORETICAL MODEL}

\label{sec1} This section aims to construct the Hamiltonian describing the
electronic properties of $\alpha-T_{3}$ bilayers. We restrict our attention
to aligned and cyclic stacking configurations. Prior to detailing the $%
\alpha-T_{3}$ bilayer, we examine the $\alpha-T_{3}$ monolayer, an extension
of the honeycomb lattice. This lattice comprises two sublattices (A and B)
augmented by additional sites (sublattice C) at the center of each hexagonal
unit cell. Within this lattice, a quasiparticle can transition between C
sites and alternate sites (e.g., B). Atoms within the A and C sublattices
occupy edge positions, each connected to three neighbouring atoms. In
contrast, B sites adopt a central hub configuration with six neighbouring
connections.The hopping amplitude between A and B sites is given by $\cos
\phi $, while that between B and C sites is $\sin \phi$, where $\phi$
dictates the hopping amplitude. This angle $\phi$ is related to another
parameter $\alpha$ via the relation $\alpha=\tan \phi$. The real parameter $%
\alpha$ serves as an interpolation between the limiting cases of the model:
the graphene lattice ($\alpha=0$) and the cubic lattice ($\alpha=1$). The
lattice is defined as an $\alpha-T_{3}$ lattice for intermediate values of $%
\alpha$ in the range $0 <\alpha<1$.
\par
\begin{figure}[H]
\centering
\includegraphics[width=1\linewidth]{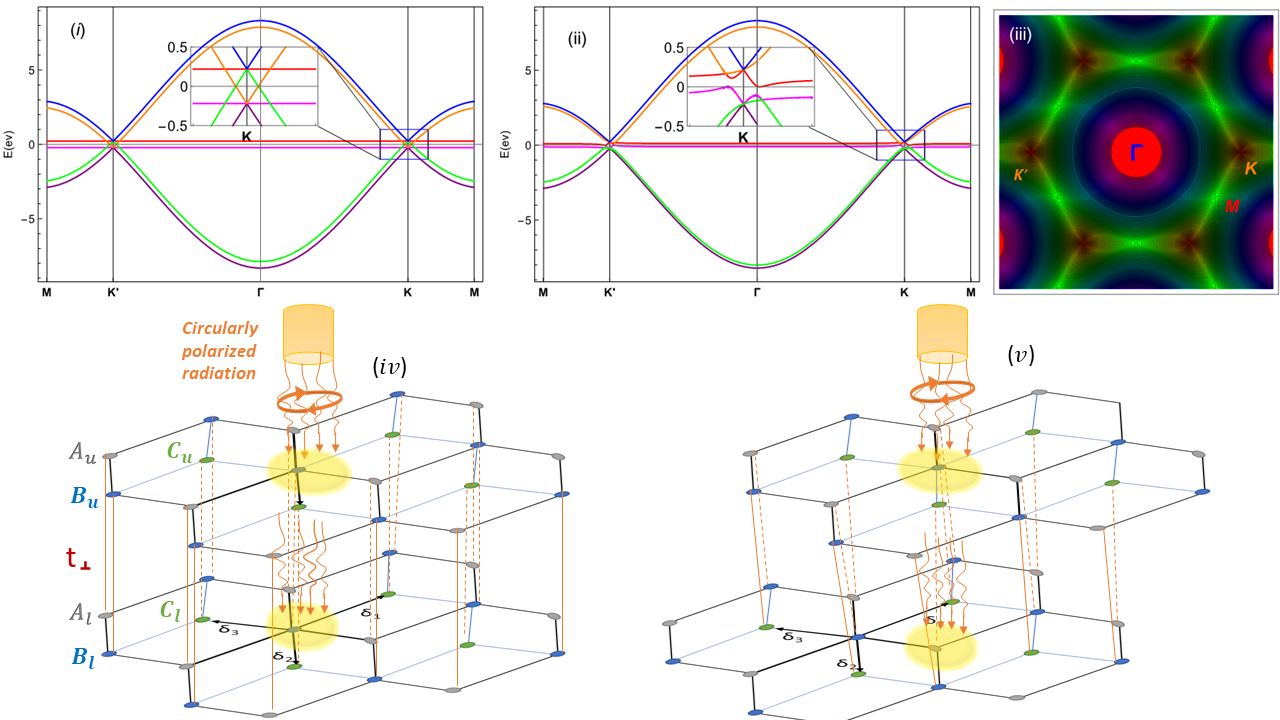}
\caption{We present the band structures of the $\protect \alpha-T_{3}$
bilayer lattice for two types of stacking: (i) aligned stacking $%
(A_{l}A_{u}-B_{l}B_{u}-C_{l}C_{u})$ and(ii) cyclic stacking $%
(A_{l}B_{u}-B_{l}C_{u}-C_{l}A_{u})$. The graph illustrates the calculated
bands along the qx axis at the high-symmetry points $(M-K^{\prime }-\varGamma%
-K-M)$ in the Brillouin zone, as shown in Figure (iii). For the
nearest-neighbor hopping, a parameter value of $t = 2.7$ eV and interlayer
coupling $t_{\perp}=t^{c}_{\perp}=t^{a}_{\perp} = 0.22$ eV are considered,
assuming these values are analogous to those of graphene\protect \cite{s37} and $(iv)$ a Schematic diagram of the aligned stacked bilayer exposed to off-resonance circular polarization, and ($v$) a schematic diagram of the cyclic stacked bilayer.}
\label{fig0}
\end{figure}

 In an $\alpha-T_{3}$ bilayer system, the $\alpha-T_{3}$ bilayer consists of
four vertically aligned, incommensurate stacking configurations \cite{HA1}.
For the purpose of this study, we will focus on only two of these stacking
arrangements. There are multiple ways to stack two proportional $%
\alpha-T_{3} $ lattices with vertically aligned sites. The simplest
configuration involves aligning corresponding sublattices in both layers: $(A_u, B_u, C_u)$ in the top layer directly above $(A_l, B_l, C_l)$ in the
bottom layer, termed 'aligned' stacking $(A_lA_u-B_lB_u-C_lC_u)$ , as shown in Figure\ref{fig0}(iv). A 'cyclic' stacking arrangement can be derived by displacing one layer of the aligned stack by a distance $a$, the lattice constant. This shift aligns sublattices $A_l$, $B_l $, and $C_l$ from one layer with $C_u$, $A_u$, and $ B_u$, respectively, in the other layer, resulting in a periodic sublattice arrangement $(A_lB_u-B_lC_u-C_lA_u)$ see Figure \ref{fig0}(v).

Within the $\alpha-T_{3}$ bilayer system there are two distinct, vertically
aligned and non-equivalent stacking configurations: $A_l A_u-B_lB_u-C_lC_u$
(aligned) and $A_l B_u-B_l C_u-C_l A_u$ (cyclic). The interlayer hopping
within these stacks is modelled by means of a tight-binding Hamiltonian \cite%
{HA2}, which includes the corresponding states of the sub-lattices $A_u$, $%
C_u$, $B_u$, $A_l$, $B_l$ and $C_l$.
\begin{align}  \label{eq1}
\mathcal{H}^{\xi}(\vect q,t^{a}_\perp,t^{c}_\perp)=%
\begin{pmatrix}
0 & \sin \phi f^{\xi }_{\vect q} & 0 & t^{a}_\perp & 0 & t^{c}_\perp \\
\sin \phi f^{\xi^{\ast} }_{\vect q} & 0 & \cos \phi f^{\xi}(q) & t^{c}_\perp
& t^{a}_\perp & 0 \\
0 & \cos \phi f^{\xi^{\ast} }_{\vect q} & 0 & 0 & t^{c}_\perp & t^{a}_\perp
\\
t^{a}_\perp & t^{c}_\perp & 0 & 0 & \sin \phi f^{\xi }_{\vect q} & 0 \\
0 & t^{a}_\perp & t^{c}_\perp & \sin \phi f^{\xi^{\ast} }_{\vect q} & 0 &
\cos \phi f^{\xi }_{\vect q} \\
t^{c}_\perp & 0 & t^{a}_\perp & 0 & \cos \phi f^{\xi^{\ast} }_{\vect q} & 0%
\end{pmatrix}%
\end{align}
The Hamiltonian $\mathcal{H}^{\xi}(\vect q,t^{a}_\perp,0)$ and $\mathcal{H}%
^{\xi}(\vect q,0,t^{c}_\perp)$ describe the electronic properties of the
aligned and cyclic $\alpha-T_{3}$ bilayer lattice, respectively. Here, $%
\vect q = (q_{x}, q_{y})$ is the wavevector in the Brillouin zone, and $f(q)
= \sum_{j=1}^{3}e^{-i\vect q·\vect {\delta_j}}$, where the vectors $%
\vect{\delta_1}$, $\vect{\delta_2}$ and $\vect{\delta_3}$ define the
relative positions of the A sites with respect to their three nearest C
neighbors. The vectors $-\vect{\delta_1}$, $-\vect{\delta_2}$ and $-%
\vect{\delta_3}$ correspond to the relative positions of the A sites with
respect to their three nearest B neighbors. The terms t represent the
nearest-neighbor hopping amplitudes. Explicitly, the vectors $\vect{\delta_j}
$ are given by: $\vect{\delta_{1}}=a( 0,1)$, $\vect{\delta_{2}}=a(\sqrt{3}%
,-1)/2$, $\vect{\delta_{3}}=a(-\sqrt{3},-1)/2$, where $a$ is the distance
between a site and its nearest neighbors, and $t_{\perp}^{a}$ and $%
t_{\perp}^{c}$ represent the interlayer coupling amplitudes for aligned and
cyclic stacking, respectively. The energy spectrum of the system, calculated
over the entire Brillouin zone, exhibits six bands. For an aligned stacking,
two bands $\varepsilon_{2,5}=\pm t_{\perp}$ are flat while the remaining
four $\varepsilon_{1,3,4,6}=\pm t_{\perp}+\mp|f(q)|$ exhibit dispersion and
electron-hole symmetry. This spectrum is independent of the parameter $%
\alpha $ and is similar to that of a bilayer graphene A-A stacking. Figure %
\ref{fig0}(i) illustrates this band structure, showing two shifted replicas
of the single-layer $\alpha-T_{3}$ band structure, with gapless band
crossings at the K (K') points. For a cyclic stacking, we have numerically
calculated the eigenvalues of the tight-binding Hamiltonian. Figure \ref%
{fig0}(ii) a presents the energy spectrum in the vicinity of the K(K')
points. We observe four dispersive bands and two flat bands that become
corrugated due to interlayer coupling. This spectrum exhibits a dependence
on the parameter $\alpha$, which we have fixed at a value of 1 in this
study. At the band crossings at the K(K') points, we note the absence of an
energy gap.\newline
In the low energy regime, where $|f(k)| << t_c$, quasiparticles exhibit
semi-metallic behaviour. The electron and hole bands occupy different of the
Brillouin zone, as shown in Figure \ref{fig0}. In the low-energy regime near
the band crossing point $t_c$ of the Dirac cone K(K'), the system behaves as
a semimetal in the absence of external fields. Introducing a Haldane gap at
the Dirac points, particularly at the crossing points $t_c$, through
external perturbations, drives the system into a topological insulator
phase. Furthermore, we demonstrate that off-resonant circularly polarized
light irradiation induces gap openings, leading to the formation of
topological states and the emergence of characteristic topological
signatures in the bilayer $\alpha-T_{3}$ lattice.

In the low energy limit, a Taylor expansion of $f(q)$ around the Dirac point
K in the Brillouin zone yields the approximation $f(k) \approx \hbar
v_{f}(\xi k_{x}-ik_{y})$. Here $k = q - K_{\xi}$ represents the wave vector
measured relative to the points $K (\xi =+1)$ and $K^{\prime }(\xi=-1)$
located at $\vect K_{\xi}=\xi4\pi/(3\sqrt{3}a)\left \lbrace
1,0\right
\rbrace $. where $v_{F}=3\gamma_{0} a/(2\hbar)$ is the Fermi
velocity and $\xi =\pm $ is the valley index.
\par We consider an $\alpha-T_{3}$ bilayer lattice being irradiated with circularly polarized radiation perpendicular to the lattice plane, as depicted in Figure \ref{fig0}($iv$)-($v$). The
associated vector potential is $\vect A(t)=A_{0}\left(\cos \omega t,\sin
\omega t\right) $, with $A_{0}=E_{0}/\omega$ and $\omega$ representing the
electric field's amplitude and the radiation's frequency, respectively.
Pierl's substitution of $\vect k\rightarrow \vect k +e \vect A(t)/\hbar$ in
equation (\ref{eq1}) leads to a modification of the Hamiltonian due to the
minimal coupling between the external radiation and the system.
\begin{equation}
\mathcal{H}^{\xi}(\vect k,t^{a}_\perp,t^{c}_\perp,t)=\mathcal{H}^{\xi}(\vect %
k,t^{a}_\perp,t^{c}_\perp)+\xi c_{0}P_{-}e^{i \xi \omega t} + \xi
c_{0}P_{+}e^{-i \xi \omega t}.
\end{equation}
where $c_{0}$ is the interaction coefficient, which is equal to $v_{F}eA_{0}$%
, $P_{-}=\mathbb{1}_{2\times2}\otimes \sigma_{-}$ and $P_{+}$ is the
conjugate of $P_{-}$, we define $\sigma_{-}$ as follows:
\begin{equation}
\sigma_{-}=%
\begin{pmatrix}
0 & 0 & 0 \\
\sin \phi & 0 & 0 \\
0 & \cos \phi & 0%
\end{pmatrix}%
.
\end{equation}
Thus the wave functions for Dirac electrons obey the time-dependent Schr%
ôdinger equation.
\begin{equation}  \label{eq2}
\mathcal{H}^{\xi}(\vect k,t^{a}_\perp,t^{c}_\perp,t) \ket{\psi(t)}=i\hbar
\partial_{t}\ket{\psi(t)}.
\end{equation}
The Hamiltonian we are currently working on is periodic $\mathcal{H}^{\xi}(%
\vect k,t^{a}_\perp,t^{c}_\perp,t+T)=\mathcal{H}^{\xi}(\vect %
k,t^{a}_\perp,t^{c}_\perp,t)$ where $T=2\pi/w$. Perhaps the appropriate tool
for dealing with such time-periodic problems is Floquet theory. According to
Floquet's theorem \cite{s1,s11,s12}, eq (\ref{eq2}) has solution of the form
of $\ket{\varphi(t)}=e^{-i\varepsilon t/\hbar }\ket{\phi(t)}$ called the
Floquet state, where $\varepsilon_{\beta}$ stands for the quasienergy and $%
|\phi(t)\big>$=$|\phi(t+T)\big>$ is periodic in time. The substitution of $%
|\varphi(t)\big>$ in the Schrödinger equation leads us to
\begin{equation}  \label{eq5}
H_{F} \ket{\phi(t)}=\varepsilon \ket{\phi(t)}.
\end{equation}
where $H_{F}=\mathcal{H}^{\xi}(\vect k,t^{a}_\perp,t^{c}_\perp,t)-i\hbar%
\partial_{t}$ is the Floquet Hamiltonian. We have a periodic Floquet state $%
\ket{\phi(t)}$ that we can write as $\ket{\phi(t)}=\sum_{m=-\infty}^{+%
\infty}e^{imwt}\ket{\phi_{m}}$ and $\mathcal{H}^{\xi}(\vect %
k,t^{a}_\perp,t^{c}_\perp,t)=\sum_{n=-\infty}^{+\infty}e^{inwt}H_{n}$ using
the Fourier transform. If we apply $H_{F}$ to the $|\phi_{n}\big>$ basis,
then we express the operator $H_{F}$ in terms of the infinite-dimensional
Hilbert-Floquet space $\mathcal{F}=\mathcal{H}\otimes \mathcal{T}$, It is
defined by the direct product of the Hilbert space $\mathcal{H}$ in a static
system and the space $\mathcal{T}$ defined by a complete set of
time-periodic functions\cite{s1,s12,s13,s14,s15}. Therefore, we can write (%
\ref{eq5}) as
\begin{equation}  \label{eq3}
\sum_{m=-\infty}^{+\infty}\left[ H_{n-m}+m\hbar \omega \delta_{m,n}\right]%
\ket{\phi_{m}}=\varepsilon_{\alpha}\ket{\phi_{n}}.
\end{equation}
We wish to express the quasi-energies as eigenvalues of the Floquet
Hamiltonian. This is not straightforward to calculate since we have an
infinity matrix equation. For this reason, we resort to describing the
formulation of an effective Hamiltonian in the high-frequency driving limit.
We assume a driving potential that is periodic in time, with a frequency $%
\omega$ that is significantly greater than the standard energy scale $%
\epsilon$ ( or the band width ) of the Hamiltonian $\mathcal{H}^{\xi}(\vect %
k,t)$, which enables us to specify a tiny paramete.
\begin{equation}  \label{eq4}
\lambda=\dfrac{\epsilon}{\hbar \omega}\ll 1.
\end{equation}
This state strategically avoids resonances (off-resonances) between the
eigenvalues of the Hamiltonian. Within this framework, light refrains from
directly exciting the electrons, opting instead to modulate the states
through the emission and re-absorption of virtual photons. As a result, we
are able to derive an effective, time-independent Hamiltonian that aptly
characterises the system.
\par To obtain the expression for the effective Hamiltonian, we utilize the
off-resonance condition derived from eq (\ref{eq4}), which specifies that $%
\varepsilon<<h w$. From this condition and the eigenvalue problem (\ref{eq3}%
), we can rewrite the zero, one, and two Fourier components of the Floquet
state with the following expression:
\begin{align}  
\label{eq8} \phi_{0}& =\dfrac{1}{\varepsilon}\left(
             H_{0}\phi_{0}+H_{1}\phi_{-1}+H_{-1}\phi_{1}\right), \\
\label{eq9} \phi_{\pm1}&=\mp \dfrac{1}{\hbar \omega}\left( H_{\pm
            1}\phi_{0}+H_{0}\phi_{\pm 1}+H_{\mp 1}\phi_{\pm 2}\right) ,\\
\label{eq10} \phi_{\pm 2}&=\mp \dfrac{1}{2\hbar \omega}\left( H_{\pm 1}\phi_{\pm
            1}+H_{0}\phi_{\pm 2}+H_{\mp 1}\phi_{\pm 3}\right).
\end{align}
The Hamiltonian Fourier modes are given by
\begin{align}
\begin{split}  \label{eqd11}
H_{s}&=\dfrac{1}{T}\int_{0}^{T} \mathcal{H}^{\xi}(\vect k,t^{a}_\perp,t^{c}_%
\perp,t)e^{-is\omega t}, \\
&=\mathcal{H}^{\xi}(\vect k,t^{a}_\perp,t^{c}_\perp)\delta_{s,0}+\xi
c_{0}P_{-}\delta_{s,\xi}+\xi c_{0}P_{+}\delta_{s,-\xi}.
\end{split}%
\end{align}
Using equations (\ref{eq9}) and (\ref{eq10}), we obtain.
\begin{align}  \label{eq11}
\phi_{\pm1}&=\mp \dfrac{\lambda}{\epsilon}\left( H_{\pm
1}\phi_{0}+H_{0}\phi_{\pm 1}\right) +\mp \dfrac{\lambda^{2}}{\epsilon^{2}}%
H_{\mp 1}\left( H_{\pm 1}\phi_{\pm 1}+H_{\mp 1}\phi_{\pm 3}\right).
\end{align}
An approximate expression for the eigenvalues of the zero component of the
wave function at high frequencies can be obtained by combining Eqs. (\ref%
{eq8}) and (\ref{eq11}) and keeping only the linear terms in $\lambda$
(where $\lambda^{2}\propto \frac{1}{\omega^{2}}$ is very small).
\begin{equation}
\left(H_{0} +\dfrac{1}{\hbar \omega}\left[ H_{1},H_{-1}\right]+\mathcal{O}
\left(1/\omega^{2} \right)\right) \phi_{0} =\varepsilon \phi_{0}.
\end{equation}
Finally, the time-independent effective Hamiltonian can be defined as
follows:
\begin{equation}
H_{eff}(\vect k,t^{a}_\perp,t^{c}_\perp)= H(\vect k,t^{a}_\perp,t^{c}_\perp)
+\dfrac{1}{\hbar \omega}\left[ H_{1},H_{-1}\right]+\mathcal{O}
\left(1/\omega^{2} \right).
\end{equation}
By applying the Fourier transform to the Fourier component of the
Hamiltonian $\mathcal{H}^{\xi}(\vect k,t)$ from eq. (\ref{eqd11}), we obtain
\begin{equation}
\dfrac{1}{\hbar \omega}\left[ H_{1},H_{-1}\right]=\Delta^{\xi}\mathbb{1}%
_{2\times2}\otimes S_{z}(\phi),\qquad S_{z}(\phi)=%
\begin{pmatrix}
-\sin^{2}\phi & 0 & 0 \\
0 & -\cos2\phi & 0 \\
0 & 0 & \cos^{2}\phi%
\end{pmatrix}.%
\end{equation}
where $\Delta^{\xi}=\xi c^{2}_{0}/\hbar \omega$. Therefore, at Dirac points,
particularly the coroissment point, light-matter coupling gives rise to a
mass term of the kind $\Delta^{\xi}\mathbb{1}_{2\times2}\otimes S_{z}$,
which lifts the six degeneracy. Virtual photon emission and absorption,
which essentially alter the static band structure, is the process this term
is responsible for.\newline
It is noteworthy that the mass term caused by off-resonance light is of the
Haldane type, which breaks time-reversal symmetry in the two valleys by
showing opposing signs. Under off-resonance consonance irradiation, the
Hamiltonian of the bilayer $\alpha-T_{3}$ can be transformed into the
Haldane model, even in the case of complicated nearest-neighbour hops
breaking time-reversal symmetry and in the absence of sublattice potentials,
as previously shown for graphene ($\alpha$ = 0) and model $\alpha-T_{3}$
monolayer in references \cite{42in,s17,84in}. 
\par When considering terms up to $\mathcal{O} \left(1/\omega \right)$, the
effective Hamiltonian (3) can be explicitly written as follows.
\begin{equation}  \label{eq16}
\mathcal{H}^{\xi}_{eff}(\vect k,t^{a}_\perp,t^{c}_\perp)=%
\begin{pmatrix}
-\Delta^{\xi}\sin^{2}\phi & \sin \phi f^{\xi }_{\vect k} & 0 & t^{a}_\perp &
0 & t^{c}_\perp \\
\sin \phi f^{\xi^{\ast} }_{\vect k} & -\Delta^{\xi}\cos2\phi & \cos \phi
f^{\xi}(k) & t^{c}_\perp & t^{a}_\perp & 0 \\
0 & \cos \phi f^{\xi^{\ast} }_{\vect k} & \Delta^{\xi}\cos^{2}\phi & 0 &
t^{c}_\perp & t^{a}_\perp \\
t^{a}_\perp & t^{c}_\perp & 0 & -\Delta^{\xi}\sin^{2} & \sin \phi f^{\xi }_{%
\vect k} & 0 \\
0 & t^{a}_\perp & t^{c}_\perp & \sin \phi f^{\xi^{\ast} }_{\vect k} &
-\Delta^{\xi}\cos2\phi & \cos \phi f^{\xi }_{\vect k} \\
t^{c}_\perp & 0 & t^{a}_\perp & 0 & \cos \phi f^{\xi^{\ast} }_{\vect k} &
\Delta^{\xi}\cos^{2}\phi%
\end{pmatrix}
.
\end{equation}
This Hamiltonian provides an important observation: It is possible to realize the $\alpha-T_{3}$ bilayer sublattice within the Haldane model, either by using a laser field or by controlling the parameter $\alpha$ such that $\alpha = 1$. In this case, we obtain the typical Haldane Hamiltonian for the $T_{3}$ bilayer lattice, where the Hamiltonian in the low-energy case is expressed as daig($\mathcal{H}^{\xi}_{eff}(\vect k,t^{a}_\perp,t^{c}_\perp$)=$(-\Delta^{\xi},0,\Delta^{\xi},-\Delta^{\xi},0,\Delta^{\xi})$ when $\Delta^{\xi}=-t_{2}3\sqrt{3}$ and a pure imaginary phase $\varphi=\pi/2$ is realized as described in ref\cite{86in}.
\par Discrete symmetries, such as charge conjugation and time reversal symmetries, are crucial in many condensed matter systems, allowing electronic states and order parameters to be classified.\newline
Time reversal symmetry is the first symmetry that we analyze. The following
is the definition of the operator that reverses time symmetry: $\hat
T_{\alpha}\mathcal{H}^{\xi}(\vect k,t^{a}_\perp,t^{c}_\perp)\hat
T_{\alpha}^{-1}=\mathcal{H}^{\xi}(-\vect k,t^{a}_\perp,t^{c}_\perp)$. The
corresponding operator, $\hat T_{\alpha} = U_{\alpha}\hat \kappa$, always
includes a matrix and the complex conjugacy operator $\hat \kappa$. We can
now check whether the effective Hamiltonians Aligned $\mathcal{H}%
^{\xi}_{eff}(\vect k,t^{a}_\perp,0)$ and Cyclic $\mathcal{H}^{\xi}_{eff}(%
\vect k,0,t^{c}_\perp)$ are invariant under the $\hat T_{\alpha}$
time-reversal operation. After completing the calculation, we find.
\begin{align} 
\label{eq17} \hat T_{0} \mathcal{H}^{\xi,\alpha=0}_{eff}(\vect k,t^{a}_\perp,0)\hat
             T_{0}^{-1} \neq \mathcal{H}^{\xi,\alpha=0}_{eff}(\vect -\vect %
             k,t^{a}_\perp,0),\quad \hat T_{1} \mathcal{H}^{\xi,\alpha=1}_{eff}(\vect %
             k,t^{a}_\perp,0)\hat T_{1}^{-1} \neq \mathcal{H}^{\xi,\alpha=1}_{eff}(\vect -%
            \vect k,t^{a}_\perp,0). \\
\label{eq17s}\hat T_{0} \mathcal{H}^{\xi,\alpha=0}_{eff}(\vect k,0,t^{c}_\perp)\hat
            T_{0}^{-1} \neq \mathcal{H}^{\xi,\alpha=0}_{eff}(\vect -\vect %
            k,0,t^{ac}_\perp),\quad \hat T_{1} \mathcal{H}^{\xi,\alpha=1}_{eff}(\vect %
            k,0,t^{c}_\perp)\hat T_{1}^{-1} \neq \mathcal{H}^{\xi,\alpha=1}_{eff}(\vect -%
            \vect k,0,t^{c}_\perp).
\end{align}
In this case, the time reversal operators in Hamiltonian Aligned $%
U_{\alpha}=0$ and $U_{\alpha}=1$ are specified as follows for the A-A
graphene bilayer and the die bilayer, respectively:
\begin{equation}
U_{\alpha = 0}=
\begin{pmatrix}
0 & 1 \\
1 & 0%
\end{pmatrix}%
\otimes
\begin{pmatrix}
1 & 0 & 0 \\
0 & 0 & -1 \\
0 & 1 & 0%
\end{pmatrix}%
, \qquad U_{\alpha = 1}=
\begin{pmatrix}
1 & 0 \\
0 & 1%
\end{pmatrix}%
\otimes
\begin{pmatrix}
0 & 0 & 1 \\
0 & -1 & 0 \\
1 & 0 & 0%
\end{pmatrix}
.
\end{equation}
The time reversal operators for the Hamiltonian Cyclic $U_{\alpha=0}$ and $%
U_{\alpha=1}$ are as follows for the A-B stacked graphene bilayer and the
AB-BC-CA stacked die bilayer, respectively:
\begin{equation}
U_{\alpha = 0}=
\begin{pmatrix}
0 & 0 & 0 & 1 & 0 & 0 \\
0 & 0 & 0 & 0 & 0 & 1 \\
0 & 0 & 0 & 0 & -1 & 0 \\
1 & 0 & 0 & 0 & 0 & 0 \\
0 & 0 & -1 & 0 & 0 & 0 \\
0 & 1 & 0 & 0 & 0 & 0%
\end{pmatrix}%
, \qquad U_{\alpha = 1}=
\begin{pmatrix}
0 & 0 & 0 & 0 & 0 & 1 \\
0 & 0 & 0 & 0 & -1 & 0 \\
0 & 0 & 0 & 1 & 0 & 0 \\
0 & 0 & -1 & 0 & 0 & 0 \\
0 & 1 & 0 & 0 & 0 & 0 \\
1 & 0 & 0 & 0 & 0 & 0%
\end{pmatrix}
.
\end{equation}
The lack of time-reversal symmetry in Hamiltonian alignment$\mathcal{H}%
^{\xi}_{eff}(\vect k,t^{a}_\perp,0)$ and Hamiltonian cyclic $\mathcal{H}%
^{\xi}_{eff}(\vect k,0,t^{c}_\perp)$ is due to circular polarization vectors
breaking the time-reversal symmetry enforced by a mass term, as demonstrated
in equations (\ref{eq17}) and (\ref{eq17s}). Cramer's theorem, which proves that the symmetry condition is broken, leads to a gap opening at valley K or K'. We will discuss this in the section on quasi-energy. 
\par Next, we will discuss charge conjugation symmetry, also known as
particle-hole symmetry (C symmetry), which has the following definition: $%
\hat C \mathcal{H}^{\xi}(\vect k,t^{a}_\perp,t^{c}_\perp) \hat C^{-1}=-%
\mathcal{H}^{\xi}(-\vect k,t^{a}_\perp,t^{c}_\perp)$. In the matrix form $%
\hat C = M_{\alpha}\hat \kappa$, the appropriate operator is defined.\newline
By applying the aforementioned operator to the effective Hamiltonians for
alignment $\mathcal{H}^{\xi}_{eff}(\vect k,t^{a}_\perp,0)$ and cyclic $%
\mathcal{H}^{\xi}_{eff}(\vect k,0,t^{c}_\perp)$ ($\alpha = 0$ and $\alpha =
1 $), it can be demonstrated that the particle-hole symmetry invariance is
expressed in the Hamiltonian as:
\begin{align} 
\label{eq19} \hat C_{0} \mathcal{H}^{\xi,\alpha=0}_{eff}(\vect k,t^{a}_\perp,0)\hat
             C_{0}^{-1} =-\mathcal{H}^{\xi,\alpha=0}_{eff}(\vect -\vect %
             k,t^{a}_\perp,0),\quad \hat C_{1} \mathcal{H}^{\xi,\alpha=1}_{eff}(\vect %
             k,t^{a}_\perp,0)\hat C_{1}^{-1} =-\mathcal{H}^{\xi,\alpha=1}_{eff}(\vect -%
             \vect k,t^{a}_\perp,0). \\
\label{eq19s}\hat C_{0} \mathcal{H}^{\xi,\alpha=0}_{eff}(\vect k,0,t^{c}_\perp)\hat
             C_{0}^{-1} =-\mathcal{H}^{\xi,\alpha=0}_{eff}(\vect -\vect %
             k,0,t^{ac}_\perp),\quad \hat C_{1} \mathcal{H}^{\xi,\alpha=1}_{eff}(\vect %
             k,0,t^{c}_\perp)\hat C_{1}^{-1} =-\mathcal{H}^{\xi,\alpha=1}_{eff}(\vect -%
             \vect k,0,t^{c}_\perp).
\end{align}
where $M_{\alpha=1}$ and $M_{\alpha=0}$ represent the operators in the
aligned effictive Hamiltonian for AA-BB-CC bilayer die and A-A bilayer
graphene, respectively.
\begin{equation}
M_{\alpha = 1}= \tau_{z}\otimes \mathbb{1}_{3\times3}, \qquad M_{\alpha =
0}=i\tau_{y}\otimes
\begin{pmatrix}
1 & 0 & 0 \\
0 & 0 & 1 \\
0 & 1 & 0%
\end{pmatrix}.%
\end{equation}
The charge conjugation operators for the cyclic Hamiltonians, $M_{\alpha=0}$
(A-B graphene) and $M_{\alpha=1}$ (AB-BC-CA dice), are given below
\begin{equation}
M_{\alpha = 1}= i\tau_{y}\otimes \mathbb{1}_{3\times3}, \qquad M_{\alpha =
0}=i\tau_{y}\otimes
\begin{pmatrix}
1 & 0 & 0 \\
0 & 0 & 1 \\
0 & 1 & 0%
\end{pmatrix}.%
\end{equation}
$\mathbb{1}_{3\times3}$, the identity matrix, along with $\tau_{y}$ and $%
\tau_{z}$, are Pauli matrices. A band of energy $\varepsilon(k)$ will have a
partner band of energy -$\varepsilon(-k)$, according to relations (\ref{eq19}%
) and (\ref{eq19s}). This symmetry proves that the six-band system consists
of a flat energy band shifted by $t_{\perp}$ with partner -$t_{\perp}$. It
is important to note that the Hamiltonian is traceless. Radiation will
therefore not affect the flat band in the dice lattice. Therefore, for all
values of $\alpha$, the sum of the energies of the bands will always be zero.%
\newline

\subsection{Quasienergy}

\label{subsec1}

We can now determine the eigenvalues $\varepsilon_{m}(k)$ of $\mathcal{H}%
^{\xi}_{eff}(\vect k,t^{a}_\perp,t^{c}_\perp)$. These represent the
structure of the quasi-energy band of off-resonance. The characteristic
equation of the eigenvalue problem for the Hamiltonian $\mathcal{H}^{\xi}(%
\vect k,0,t^{c}_\perp)$ is very complex, so we solve it numerically. Below,
we discuss the results obtained to draw the electronic structure. The
characteristic equation of the Hamiltonian $\mathcal{H}^{\xi}(\vect %
k,t^{a}_\perp,0)$ can be summarized as a reduced \textbf{\textit{depressed
cubic equation}}: $\mathbf{(\varepsilon_{m}(k)\pm t_{\perp})^{3} +
p(\varepsilon_{m}(k)\pm t_{\perp})} +\mathbf{q = 0}$. The solutions of this
equation give us the quasi-energies in the form:
\begin{equation}
\varepsilon_{m}^{s}(k)= st_{\perp}+2\sqrt{\frac{-p}{3}}\cos \left( \dfrac{1}{%
3}\arccos \left( \frac{3q}{2p}\sqrt{\frac{3}{-p}}\right) -\dfrac{2 \pi m}{3}%
\right).
\end{equation}
where
\begin{align}
p&=-\left[ \mid f_{\vect k}^{\xi}\mid^{2}+\dfrac{\Delta^{\xi ^{2}}}{8}\left(
5+3\cos 4\phi \right) \right],\qquad q=-\dfrac{\Delta^{\xi ^{2}}}{8}\sin
2\phi \sin 4\phi.
\end{align}
We consider two cases: ($m=0,1,2$ and $s=1$) and ($m=0,1,2$ and $s=-1$),
which represent the quasi-energies of the conduction and valence bands,
respectively. Due to interlayer coupling, these quasi-energies exhibit a
shift from a single point near $t_{\perp}=0$ to two points located close to
the band crossing points at $st_{\perp}$. This phenomenon leads to the
formation of a flat band in the conduction band at $t_{\perp}$ and another
in the valence band at -$t_{\perp}$. Off-resonance radiation significantly
modifies the band structure of the undriven $\alpha-T_{3}$ bilayer lattice.
It depends on both $\alpha$ and d near the Dirac point and specifically at
the band crossing points. When the parameter $\alpha$ is varied, interesting
physical properties are observed, which will be investigated further. For
clarity and consistency, I propose using the following simplified notation
for quasi-particle energies in subsequent calculations: $\varepsilon_{1}(k)=
\varepsilon_{2}^{-1}(k)$, $\varepsilon_{2}(k)= \varepsilon_{1}^{-1}(k)$, $%
\varepsilon_{3}(k)= \varepsilon_{0}^{-1}(k)$, $\varepsilon_{4}(k)=
\varepsilon_{2}^{1}(k)$, $\varepsilon_{5}(k)= \varepsilon_{1}^{1}(k)$, and $%
\varepsilon_{6}(k)= \varepsilon_{1}^{0}(k)$.
\begin{figure}[H]
	\centering
	\includegraphics[width=1\linewidth]{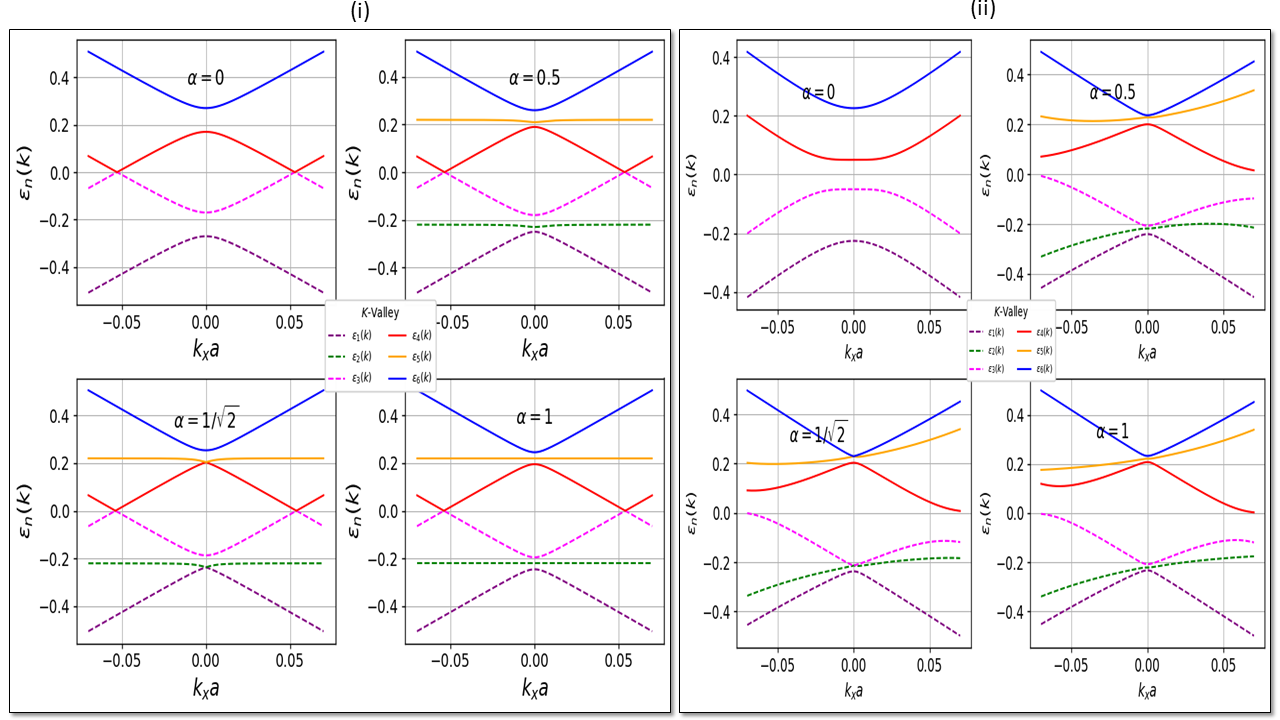}
	\caption{The influence of the $\alpha$ parameter on the quasi-energy dispersion in the K-valley of an aligned stack (i) and a cyclic stack (ii) is illustrated by the graph presented. The results obtained highlight the effect of this parameter on dispersion for different values of $\alpha$, considering circularly polarized light with an energy of $\Delta = 50 meV$}.
	\label{fig2}
\end{figure}
In this paragraph, we explore the evolution of topological band structures
by varying $\alpha$ continuously. We will analyze the band structures of an $\alpha-T_{3}$ bilayer with aligned and cyclic stacking, illustrated in Figures \ref{fig2}(i) and \ref{fig2}(ii). The choice of the coupling strengths
between the different sublattices of the two layers seems to influence the
energy spectrum of the $\alpha-T_{3}$ bilayer lattice. In the aligned stack,
the bands are dispersed and flat, while the coupling between the layers in
the cyclic stack induces waviness in the flat bands. We consider the system
to be irradiated with right-circularly polarized radiation. The amplitude
and frequency are chosen such that $\Delta = 50$ meV. Equation (\ref{eq16})
provides a description of the two non-equivalent incommensurate stackings,
both for the case with and without the inclusion of circular polarization.%
\par In bilayer geometry, the effect of circularly polarized radiation on the
band structure of the system manifests itself in two primary ways for the
two stacking configurations of the $\alpha-T_{3}$ bilayer lattice, at $%
\alpha = 0$ and $\alpha =1$. Firstly, it leads to the breaking of
time-reversal symmetry, which in turn results in the complete lifting of the
triple degeneracy of the valence and conduction bands at the Dirac point ($k
= 0$) through the opening of a gap. However, the scenario corresponding to $%
\alpha = 1/\sqrt{2}$ exhibits distinct behavior. Here, the valence and
conduction bands exhibit a flat and corrugated band topology in both the
aligned and cyclic stackings. Notably, a topological phase transition occurs
at $\alpha = 1/\sqrt{2}$, as previously observed in an $\alpha-T_{3}$
monolayer \cite{s17}. Secondly, the circular radiation deforms the flat and
corrugated band near the Dirac point ($k = 0$) for intermediate values of $%
\alpha$ ($0< \alpha <1$). Interestingly, the particle-hole symmetry is
broken in both the aligned and cyclic stackings. However, this symmetry
remains preserved for graphene ($\alpha = 0$) and the dice lattice ($\alpha
= 1$).\newline
For instance, in an irradiated dice lattice with aligned stacking, the
energy spectrum is described by the equation: $st \pm( |f^{\xi }_{\vect %
k}|^{2} + \Delta^{2})^{1/2)}$, where $st$. Notably, this equation
demonstrates that the "flatness" of the cyclic stacking's characteristic
flat and wavy band is preserved even under high-frequency radiation. It is
important to note the particle-hole transformation, where the energy of an
electron $\varepsilon(k)$ is replaced by the energy of a hole $%
-\varepsilon(-k)$.

Circular polarization breaks the system's time-reversal symmetry, causing
the six bands to split and leading to the creation of a finite zero-energy
gap. The alpha parameter determines whether this gap is closed in the
conduction and valence bands. This closure can be derived by obtaining the
eigenvalues at a Dirac point ($k = 0$) in the aligned stacking. The formula
for the gap is $\delta=\Delta/2(1-3\cos \phi)$, as shown in reference \cite%
{86in}. Including a Haldane flux instead of circular polarization preserves
the insulating state in the cyclic stacking. In the absence of radiation,
Kramer's degeneracy is ensured by time-reversal symmetry , regardless of the
alpha value. However, introducing radiation, which breaks time-reversal
symmetry, lifts this degeneracy.
\begin{figure}[H]
	\centering
	\includegraphics[width=1\linewidth]{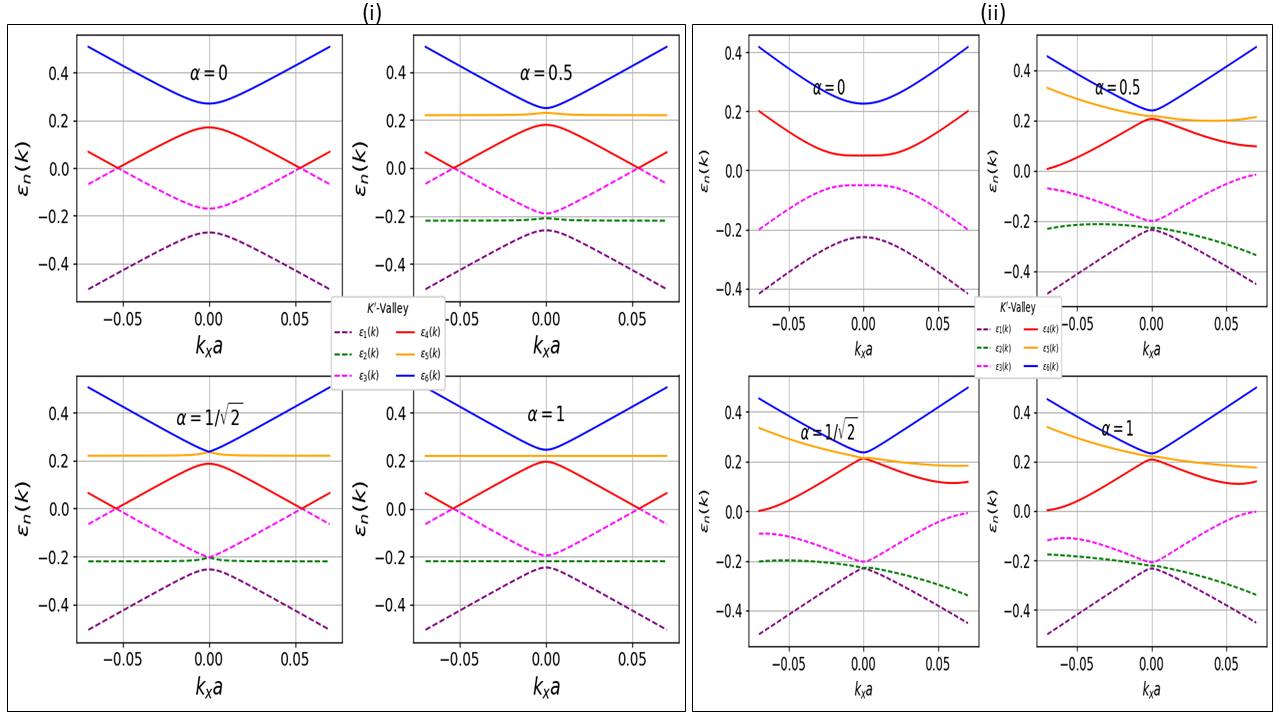}
	\caption{We repeat figure \protect \ref{fig2} for the K'-valley}
	\label{fig3}
\end{figure}
Figures \ref{fig3}(i) and \ref{fig3}(ii) show the quasiparticle energy dispersion
for the K' valley in both aligned and cyclic stacking configurations, using
the same parameters. Notably, the conduction and valence bands below the
flat band touch the deformed flat band at $\alpha = 1/\sqrt{2}$.
Interestingly, the behavior observed in the K' valley can be replicated in the K valley by simply inverting the radiation polarization $\Delta$ to -$\Delta$ for the aligned stack.\newline
The band topology remains unchanged if the energy gap of the conduction and
valence bands does not close and reopen during continuous adjustment of the
parameter. In this case, we observed that one of the gaps closes at $\alpha = 1/\sqrt{2}$ and reopens at $\alpha \neq 1/\sqrt{2}$. Therefore, we
anticipate a band topology transition at $\alpha = 1/\sqrt{2}$. In this context, a change in Berry curvature and orbital magnetic moment can be used to identify whether the system undergoes a topological transition.
\subsection{ Berry Curvature}\label{subsec2}
 In quantum systems, particularly those with Bloch bands,
Berry curvature emerges as a significant geometric property. This curvature
plays a crucial role in various physical phenomena, especially around a
Dirac point, including the anomalous quantum Hall effect and topological
transport effects. When considering an adiabatic system –where
changes occur slowly enough for the quantum state to continuously adapt
–the electron traversing k-space around the Dirac point
accumulates a geometric phase in addition to the dynamical phase. This
geometric phase is intimately linked to the concepts of Berry connection and
Berry curvature. In momentum space, the Berry connection acts as a vector
potential, denoted by $\vect A_{m}^{\xi}(\vect k)=i\bra{u_{m}^{\xi}(\vect k)}
\nabla_{k}\ket{u_{m}^{\xi}(\vect k)}$. Here, $\ket{u_{m}^{\xi}(\vect k)}%
=e^{-i\vect k\vect r}\ket{\phi_{m}^{\xi}(\vect k)}$ represents the periodic
part of the nth occupied band's Bloch function in valley $\xi$. The Berry
curvature, particularly its $z$ component denoted by $\Omega_{m}^{\xi}(\vect %
k)=(\nabla_{k}\times \vect A_{m}^{\xi}(\vect k))_{z}$, is derived from this
connection. This $z$ component, analogous to the magnetic field in real
space, exhibits gauge invariance. More specifically, the $z$ component of
the Berry curvature has been calculated numerically (as detailed in \cite%
{s20}) and can be expressed as follows.
\begin{equation}  \label{eqb1}
\Omega_{m}^{\xi}(\vect k)=-2\hbar^{2}\mathfrak{Im}\sum_{m^{\prime }\neq m}%
\dfrac{\bra{u_{m}^{\xi}(\vect k)}v_{x}\ket{u_{m'}^{\xi}(\vect k)}%
\bra{u_{m'}^{\xi}(\vect k)}v_{y}\ket{u_{m}^{\xi}(\vect k)}}{%
(\varepsilon_{m}^{\xi}(\vect k)-\varepsilon_{m^{\prime }}^{\xi}(\vect k))^{2}%
},
\end{equation}
The summation extends over all $n$ occupied bands, with the term $%
v_{i}=\hbar^{-1}\nabla_{k_{i}}\mathcal{H}^{\xi}{eff}(\vect{k},
t^{a}_{\perp}, t^{c}_\perp)$ representing the effective velocity operator in
a specific direction $(i = x, y)$. In essence, Berry curvature describes how
an electron's wave function responds to the curvature of its energy band as
its momentum varies within a material. However, as shown in Eq. (\ref{eqb1}%
), the Berry curvature becomes singular when the energy levels of two or
more states are degenerate. In systems exhibiting symmetries under spatial
inversion ($\mathcal{I}$), particle-hole transformation ($\mathcal{C}$), and
time reversal (T), the Berry curvature transforms according to specific
rules: $\mathcal{I}^{-1}\Omega_{m}(\vect{k})\mathcal{I}=\Omega_{m}(-\vect{k}%
) $, $\mathcal{C}^{-1}\Omega_{\bar{m}}(\vect{k})\mathcal{C}=-\Omega_{m}(-%
\vect{k})$, and $T^{-1}\Omega_{m}(\vect{k})T=-\Omega_{m}(-\vect{k})$. In
crystals that possess all three symmetries simultaneously, the Berry
curvature vanishes identically across the entire Brillouin zone. Conversely,
in systems where any of these symmetries are broken, the Berry curvature can
become significant, playing a crucial role in their electronic properties.
Where the label $\bar{m}$ indicates the band of energy -$\varepsilon_{m}(-k)$

\begin{figure}[H]
	\centering
	\includegraphics[width=1\linewidth]{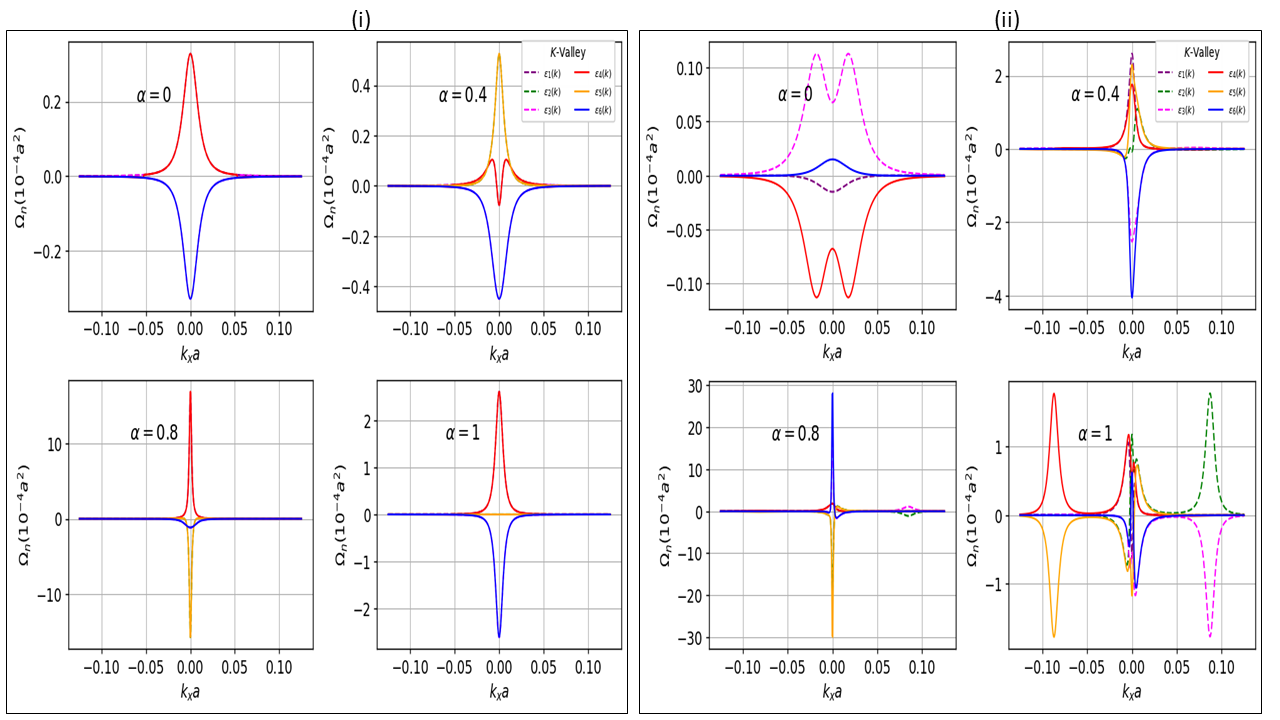}
	\caption{Berry curvature curves $\Omega_{m}^{+1}(\vect k)$ for aligned (i) and cyclic (ii) stacking near the K-valley are plotted for different values of $\alpha$, namely $\alpha=0$, $\alpha=0.4$, $\alpha=0.8$, and $\alpha=1$. We consider irradiation of the $\alpha-T_{3}$ bilayer lattice with circularly polarized light of amplitude $\Delta=50$ meV}.
	\label{fig4}
\end{figure}

\begin{figure}[H]
	\centering
	\includegraphics[width=1\linewidth]{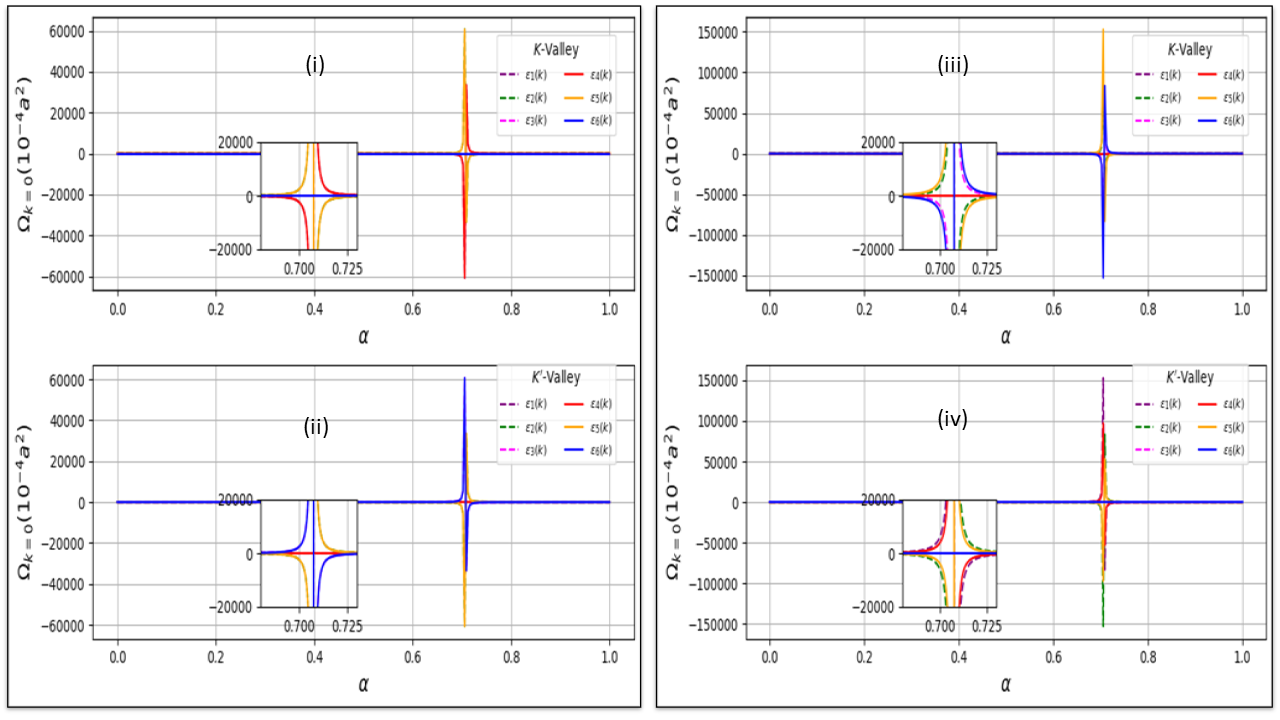}
	\caption{The maximum value of $\Omega(k=0)$ at $k=0$ is plotted as a function of $\alpha$ for (i) the K-valley and (ii) the K'-valley of the aligned stack, as well as for (iii) the K-valley and (iv) the K'-valley of the cyclic stack. Here, we consider $\Delta=50$ meV. In the K (K') valley, $\Omega(k=0)$ for the band change sign discontinuously through $\protect \alpha=1/\sqrt{2}$ while $\Omega(k=0)$ for the conduction (valence) band decreases (increases) monotonically as illustrated in the inset. The change in sign of the Berry curvature through $\protect \alpha =1/\protect \sqrt{2}$ can be seen as a topological signature}
	\label{fig5}
\end{figure}

Irradiation plays a critical role in shaping the topological properties of $%
\alpha-T_{3}$ bilayer lattices with aligned and cyclic stacking geometries.
These lattices exhibit six distinct bands with non-trivial topological
character for each stacking type. As shown in Figures \ref{fig2}(i) and \ref{fig2}(ii), these bands comprise the flat (wavy) bands $\varepsilon_{2}(k)$ and $%
\varepsilon_{5}(k)$ for aligned (cyclic) stacking, respectively, alongside
the dispersive bands $\varepsilon_{1}(k)$, $\varepsilon_{3}(k)$, $%
\varepsilon_{4}(k)$, and $\varepsilon_{6}(k)$ common to both configurations.
To investigate the system's topological evolution, we systematically varied
the $\alpha$ parameter from 0 to 1. The non-zero Berry curvature, arising
from the curvature of the bands, is responsible for inducing the topological
character. Figures \ref{fig4}(i) and \ref{fig4}(ii) illustrate the $K$-valley Berry curvature behavior for irradiated aligned and cyclically stacked $\alpha-T_{3}$ bilayers (considering a fixed $\Delta$ value of 50 meV).
Notably, the Berry curvature for individual bands becomes non-vanishing due
to the breaking of time-reversal symmetry. In all observed cases, $%
\Omega_{m}(\vect{k})$ exhibits a tendency to concentrate near the valley's
end.\newline
For aligned stacking, the Berry curvature remains invariant compared to
single-layer irradiated $\alpha-T_{3}$ lattices. This phenomenon can be
attributed to the fact that the bands of bilayer lattices simply encompass
those of the single layer, experiencing a shift mediated by the interlayer
coupling strength, $t_{\perp}$. The Berry curvature itself depends on
factors beyond just the band dispersion. Interestingly, we observe that the
positive dispersion term, which incorporates the energy bands, consistently
leads to negative Berry curvature across all $\alpha$ values. Conversely,
inferring topological characteristics from the negative dispersion term's $%
\Omega_{m}(k)$ behavior proves to be more challenging. The positive
dispersion term, however, exhibits non-monotonic behavior. At $\alpha = 0$,
it's positive and peaks at $k = 0$. As $\alpha$ increases to 0.4, a
cusp-like structure with a negative peak value emerges. When $\alpha$
reaches 0.8, the Berry curvature strengthens significantly and becomes
positive. This positive character persists even with further increases in $%
\alpha$ to 1. Notably, $\Omega_{m}(k)$ exhibits a sign change around $\alpha
= 1/\sqrt{2}$, which could potentially serve as a signature of a topological
transition.\newline
Within the range $0 < \alpha < 1$, the Berry curvature associated with the
flat band acquires a non-zero value due to the breaking of particle-hole
symmetry by the external irradiation. However, for the dice lattice ($\alpha
= 1$), the flat band's contribution to the Berry curvature at $\Omega_{m}(k)$
vanishes entirely. This is because the irradiation at this specific $\alpha$
value is incapable of breaking particle-hole symmetry. Notably, the total
Berry curvature, which is the sum of individual contributions from all bands
for a given k-value, remains zero. This phenomenon, known as local
conservation of Berry curvature, ensures that curvature gained in one band
is counterbalanced by curvature of opposite sign in another.\newline
The Berry curvature exhibits complex behavior in the cyclically stacked $%
\alpha-T_{3}$ lattice. For both the conduction and valence bands, $%
\Omega_{m}(k)$ changes sign across all $\alpha$ values, making it
challenging to directly infer topological characteristics from their
behavior. At $\alpha=0$, the Berry curvature of the valence band $%
\varepsilon_{1}(k)$ is negative, while that of the conduction band $%
\varepsilon_{4}(k)$ is positive, forming a cup-like structure at $k = 0$. It
displays non-monotonic behavior as $\alpha$ increases to 0.4. Interestingly,
for both corrugated bands $\varepsilon_{2}(k)$ and $\varepsilon_{5}(k)$, $%
\Omega_{m}(k)$ transitions from negative to positive at $\alpha= 0.4$. An
intriguing observation is the sign change of $\Omega_{m}(k)$ around $\alpha=
1/\sqrt{2}$ for bands $\varepsilon_{3}(k)$ and $\varepsilon_{1}(k)$ at $%
\alpha = 0.8$. This behavior, previously observed in aligned stacks and
single-layer $\alpha-T_{3}$ lattices, could potentially indicate a
topological signature, similar to the previously reported topological phase
transition in Haldane term bilayer cyclic $\alpha-T_{3}$ stacks \cite{s20}.
Furthermore, the corrugated bands $\varepsilon_{2}(k)$ and $%
\varepsilon_{5}(k)$ also exhibit a sign change in $\Omega_{m}(k)$ at $\alpha
= 1/\sqrt{2}$. In fact, for $\alpha = 1$, all bands display non-monotonic
behavior with sign changes. Notably, the contribution of the corrugated band
to $\Omega_{m}(k)$ remains non-zero and reaches its maximum value at $k = 0$
and $k_{x}a\approx0.085$. However, external radiation at this $\alpha$ value
fails to break the particle-hole symmetry. For the range $0 < \alpha
\leqslant 1$, the Berry curvature associated with the corrugated bands ($%
\varepsilon_{2}(k)$ and $\varepsilon_{5}(k)$) becomes non-zero, breaking the
particle-hole symmetry. This effect is absent only at $\alpha= 1$, where the
local conservation of Berry curvature is maintained due to the cancellation
of individual band contributions, resulting in a total Berry curvature of
zero for any given k-value.\newline
We investigated the Berry curvature, $\Omega_{m}(k)$, for both aligned and
cyclically stacked $\alpha-T_{3}$ bilayer lattices. For the K(K') valley,
where $\Omega_{m}(k=0)$ reaches its maximum value, we plotted $\Omega_{m}(k)$
at $k=0$ across the entire $\alpha$ range, as shown in FIG.\ref{fig5}.\newline
For aligned $\alpha-T_{3}$ bilayer lattices in the K valley, the Berry
curvatures corresponding to the flat bands ($\varepsilon_{2,5}(k)$) and the
bands with negative dispersion ($\varepsilon_{1}(k)+t<0$ et $%
\varepsilon_{4}(k)-t<0$ ) diverge as $\alpha$ approaches $1/\sqrt{2}$. In
contrast, the curvatures for the bands with positive dispersion ($\varepsilon_{3}(k) +t>0$ and $\varepsilon_{6}(k) -t>0$) remain finite, as
illustrated in FIG.\ref{fig5}(i). This divergence stems from the fact that
the flat bands and the bands with negative dispersion touch at $k=0$ when $%
\alpha=1/2$. This degeneracy at this point introduces a singularity in the
Berry curvature. Additionally, the Berry curvatures for the flat bands
(bands with negative dispersion) change sign, transitioning from positive
(negative) to negative (positive) as $\alpha$ varies around $\alpha =1/\sqrt{%
2}$, constituting a topological signature.\newline
For the K' valley, the Berry curvature remains finite for the band with
negative dispersion, while the curvature corresponding to the positive band
and the flat bands exhibits a divergence at $\alpha=1/\sqrt{2}$. This
behavior arises from the degeneracy of the bands with positive dispersion
and the flat bands at $k=0$. The Berry curvatures for the bands with
positive dispersion and the flat bands change their signs respectively
around $\alpha=1/2$, as illustrated in FIG.\ref{fig5}(ii).\newline
In the K valley of cyclically stacked $\alpha-T_{3}$ bilayer lattices, the
Berry curvatures exhibit divergences for the wavy bands ($\varepsilon_{5,2}$%
), as well as for the valence band $\varepsilon_{3}(k)$ and conduction band $%
\varepsilon_{6}(k)$ at $\alpha=1/\sqrt{2}$. In contrast, the Berry
curvatures remain finite for the valence band $\varepsilon_{1}(k)$ and
conduction band $\varepsilon_{4}(k)$, as illustrated in FIG.\ref{fig5}(iii).
These divergences at $\alpha=1/\sqrt{2}$ arise from band crossings at $k=0$.
Additionally, $\Omega(k=0)$ for the wavy bands ($\varepsilon_{2,5}(k)$), as
well as for the valence band $\varepsilon_{3}(k)$ and conduction band $%
\varepsilon_{6}(k)$, changes sign ($+$ to $-$ or $-$ to $+$) as $\alpha$
varies around $\alpha=1/2$, potentially indicating a topological signature.%
\newline
For the K' valley, the Berry curvature remains finite for the conduction
band $\varepsilon_{6}(k)$ and valence band $\varepsilon_{3}(k)$, while
exhibiting divergences at $\alpha=1/\sqrt{2}$ for the valence band $%
\varepsilon_{1}(k)$, conduction band $\varepsilon_{4}(k)$, and the wavy
bands. These divergences stem from the degeneracy of the wavy band $%
\varepsilon_{2}(k)$ and conduction band $\varepsilon_{1}(k)$ at $k=0$, as
well as the degeneracy of the wavy band $\varepsilon_{5}(k)$ and valence
band $\varepsilon_{4}(k)$. The Berry curvatures for the wavy bands $%
\varepsilon_{2}(k)$ and $\varepsilon_{5}(k)$, as well as for the conduction
band $\varepsilon_{1}(k)$ and valence band $\varepsilon_{4}(k)$, change sign
around $\alpha=1/\sqrt{2}$, as shown in FIG.\ref{fig5}(iv).
\subsection{Orbital Magnetic moment}\label{subsec3}
\begin{figure}[H]
	\centering
	\includegraphics[width=1\linewidth]{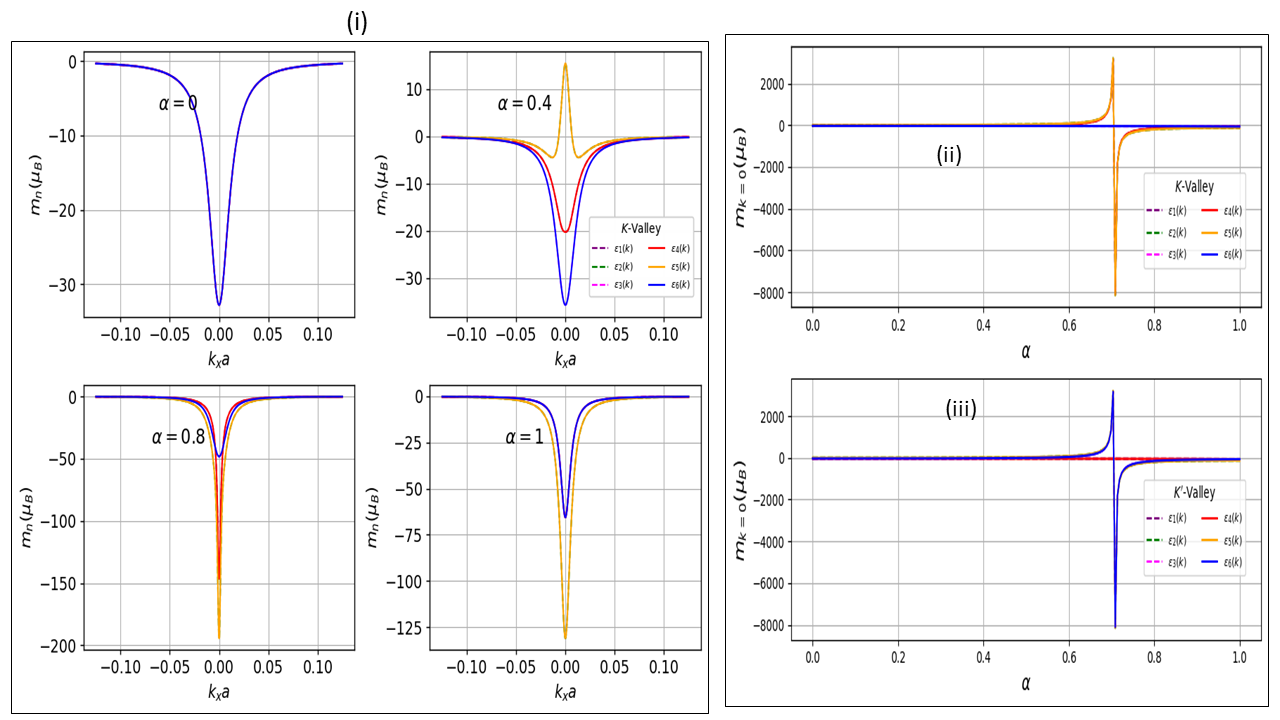}
	\caption{Variation of the orbital magnetic moment in the bilayer aligned stack  $\alpha-T_{3}$ as a function of the wave vector $k_{x}$ around the valley K: (i) for different values of the parameter $\alpha$, with $\Delta=50 $meV, based on the calculations of Eq. (\ref{eq30}); (ii) evolution of the maximum value of $m(k)$ in $k=0$ as a function of $\alpha$ in the K valley, showing discontinuities in the sign of the orbital magnetic moment associated with the flat band $\varepsilon_{2}(k)$, the valence band $\varepsilon_{1}(k)$, and the conduction band $\varepsilon_{4}(k)$ at $\alpha=1/\sqrt{2}$; (iii) the same analysis in the K' valley where the roles of the  $\varepsilon_{1}(k)$ and $\varepsilon_{4}(k)$ bands are replaced by $\varepsilon_{3}(k)$ and $\varepsilon_{6}(k)$, respectively.}
	\label{fig6}
\end{figure}

\begin{figure}[H]
	\centering
	\includegraphics[width=1\linewidth]{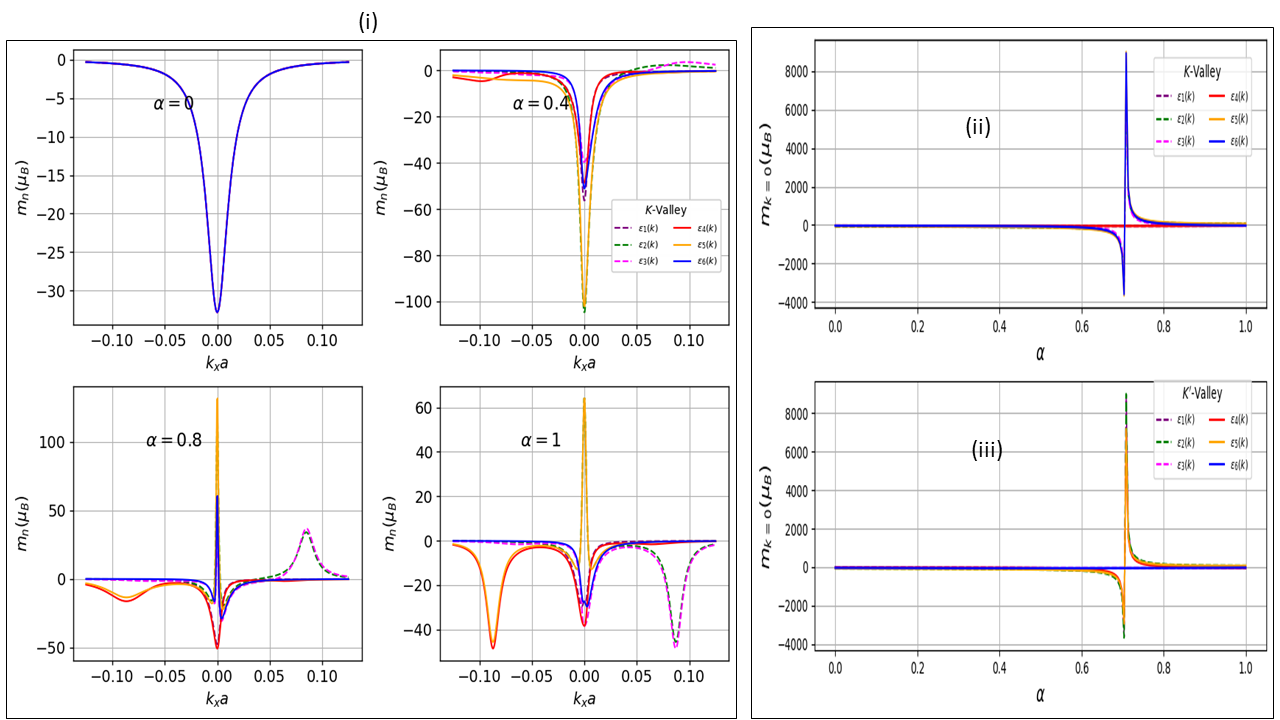}
	\caption{Orbital Magnetic Moment variation in cyclic Stacking of $\alpha-T_{3}$ bilayer as a function of $k_x$ momentum (i) around the K valley for different $\alpha$ parameters. variation of the maximum value of $m(k)$ at $k = 0$ for cyclic Stacking as a function of $\alpha$: (ii) In the K valley, with discontinuous changes in the signals from the wavy band $\varepsilon_2(k)$, band $\varepsilon_5(k)$, valence band $\varepsilon_3(k)$, and conduction band $\varepsilon_6(k)$ at $\alpha = 1/\sqrt{2}$, and (iii) In the K' Valley, where the role of the conduction band $\varepsilon_4(k)$ and valence band $\varepsilon_1(k)$ is replaced by the valence band $\varepsilon_3(k)$ for $\varepsilon_6(k)$, with $\Delta = 50$ meV.}
	\label{fig7}
\end{figure}

 A intriguing concept in condensed matter physics is the
orbital magnetic moment (OMM) associated with Bloch electrons. Bloch
electrons, represented by wave packets within Bloch bands, typically exhibit
self-rotation around their centre of mass, giving rise to an intrinsic
orbital magnetic moment. This OMM has similarities to the behaviour of
electron spin. In principle, by studying the magnetic circular dichroism
spectrum, various information can be extracted from OMM \cite{s21,s22}, so
it can be treated as a physical observable. The general expression for OMM
is given by
\begin{equation}  \label{eq28}
\vect m^{\xi}_{m}(\vect{k})=-i(e/2\hbar)\bra{\nabla_{\vect{k}}u_{m}^{\xi}(%
\vect k)}\left( H^{\xi}_{eff}(\vect{k})-\varepsilon_{m}(\vect{k})\right)%
\ket{\nabla_{\vect{k}}u_{m}^{\xi}(\vect k)},
\end{equation}
We aim to determine the simplest expression for the orbital magnetic moment
(OMM). Additionally, by differentiating the eigenvalue equation $\mathcal{H}%
^{\xi}{eff}(\vect{k})\ket{u_{m}^{\xi}(\vect k)}=\varepsilon_{m}(\vect{k})%
\ket{u_{m}^{\xi}(\vect k)} $ with respect to $\vect{k}$ and applying the
identity, we obtain
\begin{equation}  \label{eq29}
\bra{\nabla_{\vect k}u_{m'}^{\xi}(\vect k)}\ket{u_{m}^{\xi}(\vect k)}=\dfrac{%
\bra{u_{m'}^{\xi}(\vect k)}\nabla_{\vect k}\mathcal{H}^{\xi}{eff}(\vect{k})%
\ket{u_{m}^{\xi}(\vect k)}}{\varepsilon_{m^{\prime }}(\vect{k}%
)-\varepsilon_{m}(\vect{k})},
\end{equation}
Through the inclusion of the unit operator $1=\sum_{m}%
\ket{u_{m}^{\xi}(\vect
k)}\bra{u_{m}^{\xi}(\vect k)}$ and Eq.(\ref{eq29}) into Eq.(\ref{eq28}), we
can reformulate the $z$-component of Eq.(\ref{eq28}) as follows.
\begin{equation}  \label{eq30}
m_{m}^{\xi}(\vect k)=-\hbar e\mathfrak{Im}\sum_{m^{\prime }\neq m}\dfrac{%
\bra{u_{m}^{\xi}(\vect k)}v_{x}\ket{u_{m'}^{\xi}(\vect k)}%
\bra{u_{m'}^{\xi}(\vect k)}v_{y}\ket{u_{m}^{\xi}(\vect k)}}{%
\varepsilon_{m}^{\xi}(\vect k)-\varepsilon_{m^{\prime }}^{\xi}(\vect k)}.
\end{equation}
The behavior of the OMM is analyzed within the framework of Equation (\ref%
{eq30}) for irradiated $\alpha-T_{3}$ lattices with both aligned and cyclic
stacking arrangements (Figures \ref{fig6}(i) and \ref{fig7}(i) depict the OMM
distribution for the K valley). Notably, the existence of a non-zero OMM for
the Hamiltonian (\ref{eq16}) in these two stacking configurations can be
established through fundamental symmetry transformation relations of the
OMM. Under inversion $(\mathcal{I})$, particle-hole transformation $(%
\mathcal{C})$, and time-reversal (T) operations, the OMM transforms as $%
\mathcal{I}^{-1}m_{m}(k)\mathcal{I}= m_{m}(-k)$, $\mathcal{C}^{-1}m_{\bar{m}%
}(k)\mathcal{C}=m_{m}(-k)$, and $T^{-1}m_{m}(k)T = -m_{m}(-k)$,
respectively. Therefore, the existence of a non-zero $m(k)$ requires the
breaking of at least one of these symmetries. Interestingly, the OMM
distribution exhibits a peak around the valley edge $(k\approx 0)$, similar
to the Berry curvature. However, distinct features differentiate the OMM
distribution from that of the Berry curvature, suggesting the presence of
additional mechanisms influencing the OMM in irradiated $\alpha-T_{3}$
lattices.\newline
In aligned $\alpha-T_{3}$ bilayer systems, when time-reversal symmetry is
broken, it results in non-zero OMMs. Specifically, for the values $\alpha =
0 $ and $\alpha = 1$, as illustrated in Figure \ref{fig6}(i), the OMMs
associated with the conduction and valence bands are found to be negative
and coincide due to particle-hole symmetry. One fascinating observation is
that the OMM linked to non-vanishing flat bands at $\alpha = 1$ shows an
opposite sign compared to that observed in the Berry curvature. This
observation sheds light on the unique behavior of OMMs in these systems.
Within the intermediate $\alpha$ regime ($0< \alpha < 1$), the broken
time-reversal symmetry leads to distinct Orbital Magnetic Moment (OMM)
values for individual bands in $\alpha-T_{3}$ bilayers. This observation
stands in contrast to monolayer systems, where both time-reversal and
particle-hole symmetries have been observed to be broken simultaneously. We
present the variation of $m(k=0)$ with $\alpha$ for the two valleys in Fig.\ref{fig6}(ii)-(iii)  to explore the topological characteristics of the OMM. For the K
valley, the OMMs corresponding to the flat bands $\varepsilon_{2}(k)$ and $%
\varepsilon_{5}(k)$ as well as the conduction band $\varepsilon_{4}(k)$ and
the valence band $\varepsilon_{1}(k)$ change their signs respectively at $%
\alpha = 1/\sqrt{2}$. In contrast, the one associated with the conduction
band $\varepsilon_{6}(k)$ and the valence band $\varepsilon_{3}$ is monotone
with respect to $\alpha $. For the K' valley, the roles of the conduction
bands $\varepsilon_{4}(k)$ and valence $\varepsilon_{1}(k)$ are reversed,
respectively with the valence band $\varepsilon_{3}(k)$ and the conduction
band $\varepsilon_{6}(k)$.\newline
Let's now analyse the behaviour of the OMMs in the cyclic stacking of $\alpha-T_{3}$
bilayers, as mentioned above. The time inversion is broken, resulting in
non-zero OMMs, as illustrated in Fig.\ref{fig7}(i). The OMMs corresponding to
the conduction and valence bands are negative for $\alpha0$. For $\alpha = 1$%
, the OMMs for the conduction and valence bands are negative, except for the
corrugated bands, which are positive near $k_{x}\approx0$ and negative at $%
k_{x}a\approx \pm0.085$. This behaviour is not observed in aligned stacking,
which preserves particle hole symmetry. In aligned stacking the OMMs
coincide. Here the system shows a different behaviour in different cyclic
stacking than in aligned stacking, which is an unusual behaviour. For
intermediate values $(0 < \alpha < 1)$ the time-reversal, inversion and
particle-hole symmetries are broken. This leads to a fundamental difference
compared to aligned stacks. In OMMs, each individual band is assigned a
different value. This system is characterised by exotic topological
properties, as confirmed by the introduction of the Haldane term to describe
topological insulating systems\cite{86in}.\newline
Following the analogy of aligned systems, we explore the topological
characteristics of OMMs by examining the variation of $m(k=0)$ with $\alpha$
for the two valleys illustrated in Fig.\ref{fig7}(ii)-(iii). For the K valley, the
OMMs associated with the wavy bands $\varepsilon_{2}(k)$ and $%
\varepsilon_{5}(k)$, as well as the conduction band $\varepsilon_{6}(k)$ and
the valence band $\varepsilon_{3}(k)$, exhibit a sign change. Intriguingly,
we observe that the OMM transitions from a negative sign throughout the $%
\alpha$ range until it reaches $1/\sqrt{2}$, where it becomes positive. This
behavior stands in contrast to that observed in aligned stacks. In contrast,
the conduction band $\varepsilon_{6}(k)$ and the valence band $%
\varepsilon_{3}(k)$ exhibit a monotonic dependence on $\alpha$. For the K'
valley, the roles of the conduction band $\varepsilon_{4}(k)$ and the
valence band $\varepsilon_{1}(k)$ are reversed with those of the conduction
band $\varepsilon_{6}(k)$ and the valence band $\varepsilon_{3}(k)$,
respectively. The wavy band $\varepsilon_{2}(k)$ also exchanges its role
with the wavy band $\varepsilon_{5}(k)$.
\subsection{Orbital Magnetization}\label{subsec4} 
Orbital magnetization is a fascinating property of
crystalline materials where time-reversal symmetry is broken. Often referred
to as the "modern theory of orbital magnetization," this understanding is
based on developments in the Berry phase theory of orbital magnetization
\cite{s23,s24,s25}. An illustrative derivation of orbital magnetization is
presented, based on the semiclassical approach to the dynamics of wave
packets for Bloch electrons\cite{s25}, the Wannier function approach\cite%
{s24,s26}, or perturbation theory\cite{s23}. These methods address two
closely related quantities: the orbital moment magnetization (OMM) and the
Berry curvature (To understand how this phenomenon evolved from its initial
inception to its modern development, I recommend the reader to refer to this
review by T. Thonhauser\cite{s27}). Orbital magnetization is a thermodynamic
quantity that can be derived from the differentiation of the grand canonical
potential with respect to the magnetic field. This grand canonical
potential, which represents the free energy of a free particle in the
presence of a weak magnetic field $\vect B$, can be written as follows:
\begin{equation}
F^{\xi}=-\dfrac{1}{\beta}\sum_{k,m}\ln \left( 1+e^{ -\beta( E^{\xi}_{m}(%
\vect k)-\mu) } \right)
\end{equation}
In the presence of a weak magnetic field $\vect B$, the electronic energy $%
E^{\xi}_{m}(\vect k)=\varepsilon^{\xi}_{m}(\vect k)-\vect m(\vect k).\vect B$
acquires a correction term due to the orbital magnetic moment $\vect m(\vect %
k)$\cite{s28,s22}. Additionally, the parameter $\beta = 1/(k_{B}T)$, where $%
k_{B}$ is the Boltzmann constant, T is the temperature, and $\mu$ is the
chemical potential.
\par Building on a semi-classical approach, Xiao et al.\cite{xio5} rigorously
showed that the Berry curvature affects the distribution of states within
phase space. In the presence of a weak external magnetic field B, the
summation over $k$ can be transformed into an integral expressed as $%
\sum_{k}\longrightarrow \dfrac{1}{(2\pi)^{2}}\int^{\mu}(1+\dfrac{e \vect B.%
\vect \Omega^{m}(\vect k)}{\hbar})d\vect k$. This transformation arises from
the violation of Liouville's theorem due to the non-conservation of phase
space volume.
\par The magnetization is then the derivative of the free energy of the magnetic
field $\vect B$ at fixed T and $\mu$: $\vect M=-(\partial F/\partial \vect %
B)_{\mu,T}$, which leads to
\begin{equation}  \label{e32}
\vect M(\vect r) = \dfrac{1}{(2\pi)^{2}}\sum_{m} \int d\vect k f(k) \vect m(%
\vect k) + \dfrac{e}{2\pi \beta \hbar} \sum_{m}\int d\vect k \vect %
\Omega^{m}(\vect k) \log(1 + e^{-\beta(\varepsilon-\mu)})
\end{equation}
This expression for magnetization is derived under equilibrium conditions,
specifically for zero external magnetic field but applicable to a broad
range of temperatures. The term $f(\vect k)=\left( 1 + e^{\beta(\varepsilon(%
\vect k)-\mu)}\right)^{-1}$ represents the Fermi-Dirac distribution
function. The integration in equation (\ref{e32}) exhibits two different
contributions: the first term originates from the self-rotation of the wave
packet, and the second term is due to the motion of the wave packet's center
of mass.
\par
\begin{figure}[H]
	\centering
	\includegraphics[width=1\linewidth]{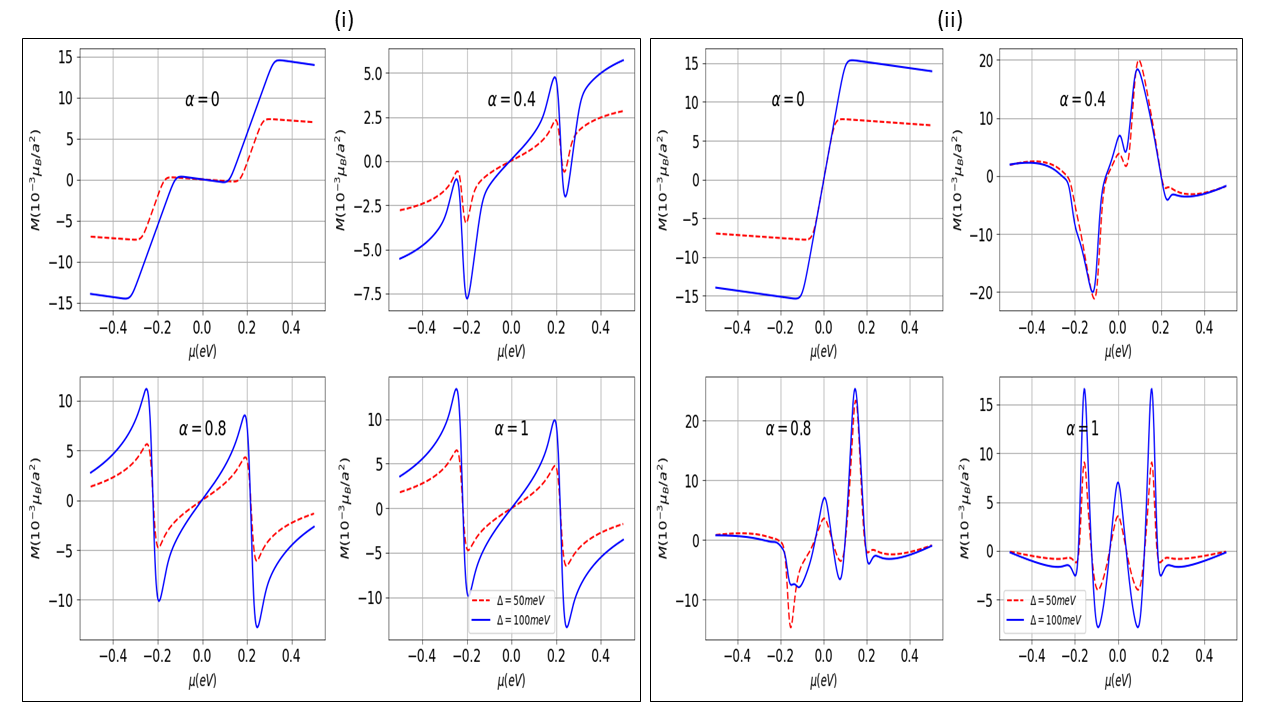}
	\caption{Orbital magnetism in an $\alpha-T_{3}$ bilayer in an aligned stack (i) and a cyclic stack (ii) at valley K: dependence as a function of chemical potential $\mu$ for different values of $\alpha$. The temperature considered is $T = 100$ K.}
	\label{fig8}
\end{figure}

\begin{figure}[H]
	\centering
	\includegraphics[width=1\linewidth]{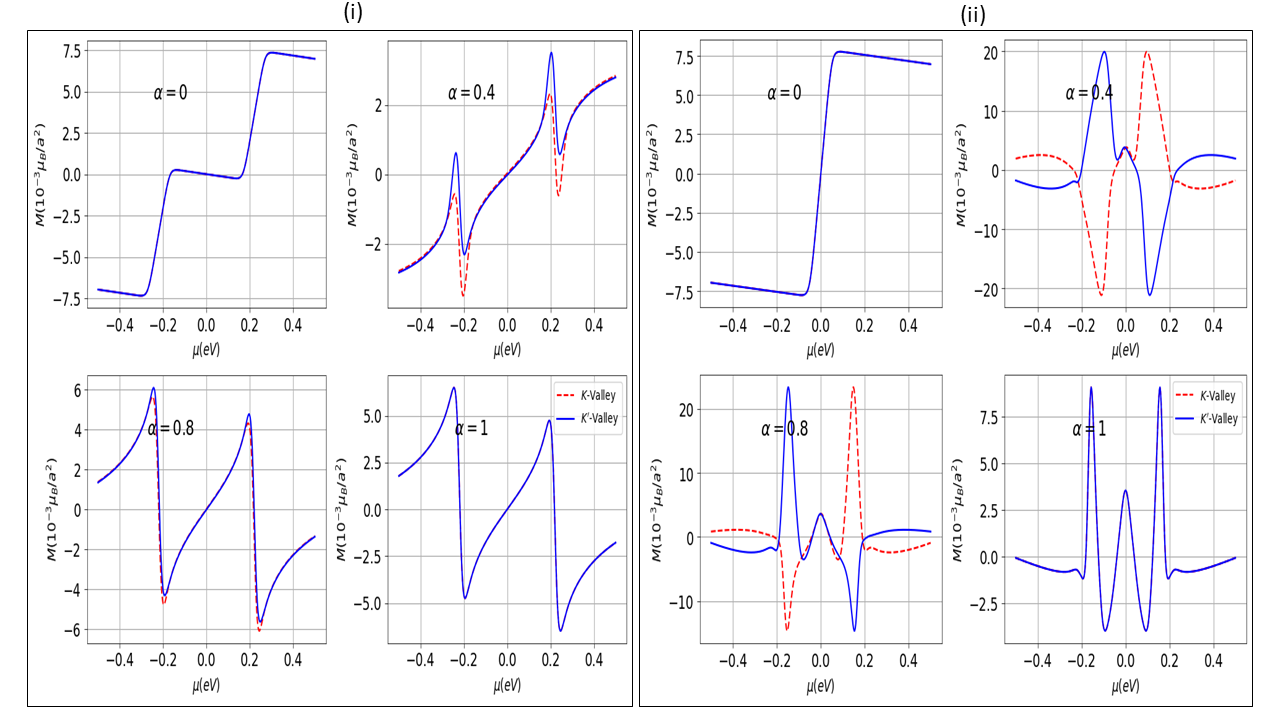}
	\caption{Distribution of Orbital Magnetization in an aligned stack (i) and a cyclic stack (ii) as a function of chemical potential $\mu$ for valleys K and K', with $\Delta=50$ meV and $T=100k$.}
	\label{fig9}
\end{figure}

 To gain insights into the behavior of $M(\vect r)$ in the two scenarios of
irradiated lattices for aligned stacking and cyclic stacking, we numerically
evaluated the integrals k in equations (\ref{e32}). To delve into the impact
of chemical potential $\mu$ on the system, we plotted the curves for
distinct values of $\alpha$ ($\alpha = 0$, 0.4, and 0.8) at a constant
temperature of $T = 100$ K, as illustrated in Figures \ref{fig8}(i) and \ref{fig8}(ii). The interactions between matter and photons are modulated by the
light-induced mass term $\Delta$. We consider two values of $\Delta$, namely $\Delta = 50$ meV and $\Delta =$ 100 meV.\newline
In aligned lattices, magnetization exhibits a significant increase with
increasing $\Delta$ for all values of $\alpha$. It displays an antisymmetric
function for $\alpha=0$ and $\alpha=1$ and changes sign with the change of
sign of $\mu$. Furthermore, it transitions from a negative (positive) value
to a positive (negative) value around $\mu=0$. Notably, the antisymmetry of
magnetization vanishes for $\alpha=0.4$ and $\alpha=0.8$, indicating its
absence in these specific cases. This disappearance can be attributed to the
breaking of the hole-particle symmetry. Topological signatures of orbital
magnetization can be visualized by examining its dependence on $\mu$ in $%
\Delta=50 meV$. For $\alpha = 0$, the magnetization exhibits a linear
variation with $\mu$, indicating that the interval of $\mu$ in which
magnetization is linear is limited by the energy gap between the conduction
bands (or valence bands). In the case of $\alpha \neq 1$, the magnetization
varies linearly with $\mu$ when the interval of $\mu$ corresponds to the
energy gap between the conduction band and the flat band (or between the
valence band and the flat band). This linear variation of magnetization with
$\mu$ in this energy gap constitutes a topological signature, as mentioned
by Ceresoli\cite{aI6}, and reveals a relationship between magnetization and
the Chern number, expressed by $dM/d\mu=e/h \sum_{m}^{occ}C_{m}$.\newline
In cyclic lattices, magnetization exhibits a significant increase with the
rise in $\Delta$ for $\alpha = 0$ and $\alpha = 1$. However, for $\alpha =
0.4$ and $\alpha = 0.8$, the magnetization does not follow the same growth
pattern. When $\alpha = 0$, the magnetization changes in an antisymmetric
manner with respect to $\mu$, while for $\alpha = 1$, it changes
symmetrically. This distinct behavior contrasts with the aligned system
where $\alpha = 1$. The symmetric and antisymmetric nature disappears for $%
\alpha = 0.4$ and $\alpha = 0.8$, due to the breaking of particle-hole
symmetry. As previously mentioned, in the aligned system, the topological
characteristics in this cyclic system cause the magnetization to vary
linearly with $\mu$, indicating that the $\mu$ interval within which the
magnetization remains linear is limited by the gap energy.\newline
We investigate the evolution of magnetization M in aligned and cyclic
stacking lattices as a function of chemical potential $\mu$, considering
both the K and K' valleys with $\Delta = 50$ meV and $T = 100 $K, as
illustrated in Figures \ref{fig9}(i) and \ref{fig9}(ii). The results show that,
due to particle-hole symmetry, the magnetization in the K and K' valleys
coincides for $\alpha = 0$ and $\alpha = 1$. However, for $0 < \alpha < 1$,
the magnetization in the two valleys exhibits a pronounced difference,
attributed to the breaking of particle-hole and time-reversal symmetries
\subsection{Thermal Hall conductivity}

\label{subsec5} 
\begin{figure}[H]
\centering
\includegraphics[width=1\linewidth]{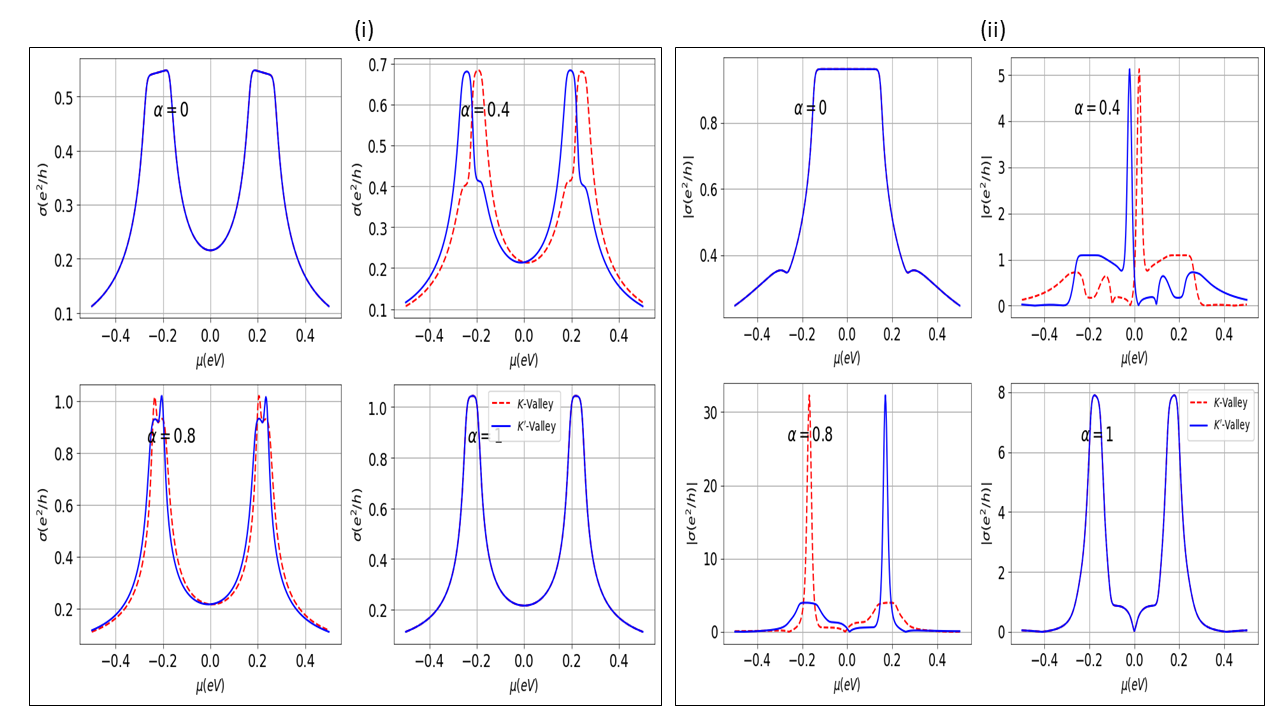}
\caption{Anomalous Hall conductivity of the aligned stack (i), with $\Delta=50 meV$, and the cyclic stack (ii), with $\Delta=100 meV$, as a function of chemical potential, for various values of $\alpha$, calculated in both K and K' valleys, at temperature $T=50k$.}
\label{fig10}
\end{figure}

This section explores the interesting field of the Anomalous
Hall Effect (AHE) in irradiated, aligned and cyclic $\alpha-T_{3}$ bilayer
lattices. The AHE has captivated researchers with its counterintuitive
nature: the spontaneous appearance of a significant Hall current in a
ferromagnetic material when exposed to an applied electric field alone,
defying the conventional requirement of an external magnetic field \cite{s30}%
. The intricate microscopic mechanisms governing AHE remain a subject of
ongoing debate within the scientific community, despite its longstanding
presence (over a century) and critical role in material characterisation
\cite{s31}.\newline
Three main mechanisms historically proposed to contribute to AHE are the
intrinsic contribution, and extrinsic contributions from obliquity and side
jump scattering \cite{s32,s33,s34,s35}. Specifically, the intrinsic
contribution, which is independent of scattering processes, is attributed to
the anomalous velocity induced by the Berry phase, a geometric property
characterizing the electronic band structure. This intrinsic contribution
can be conceptualized as an unquantized counterpart to the well-established
quantum Hall effect. Interestingly, some earlier work suggests a particularly remarkable scenario: Berry curvature alone determines anormal Hall conductivity (AHC)\cite{s36}.
\begin{equation}  \label{eq33}
\sigma_{xy}^{\xi}=\dfrac{e^{2}}{\hbar}\sum_{m}\int \dfrac{d^{2}k}{(2\pi)^{2}}
\Omega_{m}^{\xi}(\vect k)f^{\xi}_{m}(\vect k)
\end{equation}
To elucidate the intrinsic contribution to the AHE, we leverage a generic
effective Hamiltonian (\ref{eq16}) that captures the electronic properties
within the aligned and cyclically irradiated bilayer network. Notably, the
electronic bands exhibit band crossings at points denoted as $t^a$( $t^c$).
A mass term, induced by a circular vector potential, separates these bands.
Subsequently, the AHC, $\sigma_{xy}$, is calculated numerically using Eq (%
\ref{eq33}). Figures \ref{fig10}(i) and \ref{fig10}(ii) depict the evolution of the
AHC as a function of the chemical potential for the aligned and cyclic
systems, respectively. These calculations are conducted at a fixed
temperature of $T = 50$ K, with an irradiation strength of $\Delta = 50$ meV
for the aligned stacking and $\Delta = 100$ meV for the cyclic stacking.%
\newline
To investigate the intrinsic contribution to the AHC within aligned $\alpha-T_{3}$ bilayer lattices, we first examine the AHC for both valleys. As shown in Figure \ref{fig10}(i), the AHC for both valleys can be represented as a superposition of two AHC blocks originating from the single layer $\alpha-T_{3}$ systems. These blocks are shifted relative to each other as
the chemical potential $\mu$ approaches the flat energy region
(equivalently, the interlayer coupling energy, $t^{a}$) and deviates from
zero. When $\mu$ is in the gap region $\Delta$, where AHC reaches its
maximum values and approaches $\frac{1}{2} e^{2}/h(\alpha=0)$ and $%
e^{2}/h(\alpha=1)$, the states with occupied energy just below $\Delta$
contribute most to the Hall conductivity $\sigma_{xy}$. For $\mu$ values
above $\Delta$, the contributions from the upper and lower bands cancel each
other out, leading to a decrease in AHC as $\mu$ moves further away from the
gap $\Delta$. This behaviour highlights the significant enhancement in
conductivity when $\mu$ is confined within the gap region $\Delta$. The flat
bands themselves do not contribute to the Hall conductivity. Since the
corresponding Berry curvature vanishes, the flat bands themselves do not
contribute to $\sigma_{xy}$. Furthermore, it is observed that the $%
\sigma_{xy}$ for $\alpha=1$ is double that for $\alpha=0$. Since the
particle-hole symmetry is preserved, the conductivities for both valleys
coincide at $\alpha=0$ and $\alpha=1$. In contrast to the behavior observed
at $\alpha=0$ and $\alpha=1$, where particle-hole symmetry holds, the
scenario for $\alpha \neq0,1$ reveals distinct characteristics. Here, we
witness the emergence of valleys exhibiting contrasting behavior in the $%
\sigma_{xy}$. This observation signifies a breakdown of particle-hole
symmetry within the system. Similar to the cases of $\alpha=0$ and $\alpha=1$%
, two distinct AHC blocks emerge at $\mu=t^{a}$ and $\mu=-t^{a}$. Focusing
on the block where $\mu$ is near $t^{a}$, a unique two-plateau structure
becomes evident, arising from the presence of two band gaps of unequal
sizes. As $\mu$ approaches $t^{a}-\Delta$, the plateau associated with
valley K' comes into focus. Conversely, when $\mu$ approaches $t^{a}+\Delta$%
, the plateau corresponding to valley K dominates. The total AHC, which
represents the combined contributions from both valleys, exhibits a notable
behavior. It approaches the quantized values of $e^{2}/h$ and $2e^{2}/h$
within the band gaps t-d and $t^{a}+\Delta$, when $\alpha$ falls below ($%
\alpha<1/\sqrt{2}$) and exceeds ($\alpha>1/\sqrt{2}$) the value of $1/\sqrt{2%
}$. This observation provides strong evidence for a topological phase
transition occurring at a critical value of $\alpha=1/\sqrt{2}$.
This analysis explores the anomalous Hall conductivity of the cyclically
stacked $\alpha$-T$_{3}$ bilayer as a function of the chemical potential $%
\mu $, as illustrated in Figure \ref{fig10}(ii). It is observed that the Hall
conductivity $\sigma_{xy}$ for both valleys is identical, reflecting the
conservation of particle-hole symmetry for $\alpha = 0$ and $\alpha = 1$.
Within the bandgap, variations in the chemical potential result in the
formation of a plateau in $\sigma_{xy}$, with a width proportional to $%
\Delta $, which arises from the contributions of all occupied states in the
valence band. In contrast to the aligned configuration, the corrugated band
provides an additional contribution to $\sigma_{xy}$ due to its
non-vanishing Berry curvature. In $\alpha-T_{3}$ bilayers with $\alpha \neq
0 $ and $\alpha \neq1$, the breakdown of particle-hole symmetry gives rise
to valley-dependent behaviour in $\sigma_{xy}$, which is distinct from the
behaviour observed in bilayers with other values of $\alpha$. The presence
of two unequal forbidden bands in the quasi-energy spectrum for each valley
results in a two-plateau structure near the bandgap edge ($\mu \approx t^{c}$
and $\mu \approx -t^{c}$ ). These properties are more pronounced at
decreased temperatures and elevated $\Delta$ values, although they are not
included in this analysis. The total AHC, $\sigma_{xy}$, is the sum of
contributions from the K and K' valleys. As illustrated in Figure\ref{fig10}(ii)
, within the band gaps of these valleys, the AHC approaches a quantized
value of $1e^{2}/h $ for $\alpha<1/\sqrt{2}$, $4e^{2}/h$ for $\alpha=0.4$,
and $16e^{2}/h$ for $\alpha=1$. The exceptionally high AHC observed for $%
\alpha = 0.8$ and $\alpha = 1$ confirms a topological phase transition at $%
\alpha = 1/\sqrt{2}$.\newline
In the off-resonance irradiated cyclic stack and effective control of external parameters, notably the amplitude of the light field, proportional to $\Delta$. In this context, we find $\sigma_{xy}$ plateaus appearing for values of $\alpha = 0.4$ and 0.8. The responses of the two valleys, K and K', show values close to $1e^{2}/h$ and $4e^{2}/h$, similar to those obtained in the Haldane model applied to a cyclic stack, as discussed in P.Parui\cite{86in}. Thus, Floquet Engineering presents a programmable solution to realize the Haldane model and control topological phases.
\section{Conclusion}

\label{sec2} In this study, we investigated the topological signatures
induced by circularly polarized light on $\alpha -T_{3}$ bilayer networks.
The bilayer $\alpha -T_{3}$ lattice we examined exhibited two distinct
commensurate stacking configurations: AA-BB-CC alignment and AB-BC-CA
cyclic. Our effective model accurately captured the topological features
associated with the K and K' band crossing points. As we have seen, circular off-resonance polarization breaking time reversal symmetry induces a Haldane-type Chern insulator in the $\alpha-T_{3}$ bilayer when the phase $\varphi = \pi/2$ and $\Delta^{\xi} = -t_{2}3\sqrt{3}$ are fixed at the Dirac point\cite{86in}. This makes it possible to realize a Haldane insulator by controlling the system parameters and the polarization amplitude. By analyzing the discrete
symmetries of the system, we identified a broken time-reversal symmetry at $%
\alpha =0$ and a broken particle-hole symmetry at $\alpha =1$. These
stacking configurations exhibit distinct energy spectra. By analyzing
various Berry phase effects, we identified topological signatures in both
the aligned and cyclic $\alpha -T_{3}$ lattices. Our calculations of Berry
curvature, orbital magnetic moment, orbital magnetization, and anomalous
Hall conductivity revealed the topological nature of these quantities.
Notably, a critical transition occurs at $\alpha =1/\sqrt{2}$, where the
Berry curvature and orbital magnetic moment associated with the flat band in
aligned stacks and the corrugated band in cyclic stacks undergo a
significant change. The light-driven effect on the flat and corrugated bands
near the Dirac points results primarily in two distinct, $\alpha $%
-dependent, equal-width bandgaps at the $t^{a}/t^{c}$ intersections. Within
these bandgaps the orbital magnetization shows a linear dependence on the
chemical potential. The gradient of this linear relationship is steeper for $%
\alpha =0.4$ compared to $\alpha =0.8$. Around $\alpha =1/\sqrt{2}$ this
difference in steepness is strongly correlated with the Chern number
transition. As the chemical potential falls into the band gap, the anomalous
Hall conductivity takes on quantized values, i.e. a plateau. For $0<\alpha
<1 $, for aligned stacking and cyclic stacking, a 'two plateau' structure is
observed in the Hall conductivity of the K and K' valleys. Nevertheless, the
total anomalous conductivity of the individual valleys k and k' with respect
to aligned stacking in band gaps when the chemical potentials are $%
t^{a}+\Delta $ and $t^{a}-\Delta $ or $-t^{a}-\Delta $ and $-t^{a}+\Delta $
approaches $e^{2}/h$ and $2e^{2}/h$ respectively, approximately when $\alpha
<1/\sqrt{2}$ and $\alpha >1/\sqrt{2}$, respectively, in cyclic stacking the
AHC approaches a quantum value of $1e^{2}/h$ for $\alpha <1/\sqrt{2}$, $%
4e^{2}/h$ for $\alpha =0.4$ and $16e^{2}/h$ for $\alpha =1$. As we observe plateaus at $\alpha = 0.4$ and 0.8 in the K and K' valleys, these correspond to the plateaus in the Haldane model for cyclic stacking. In both stacks,
for $0<\alpha <1$, broken particle-hole symmetry leads to distinct valley
patterns in the orbital magnetisation and the AHC. By analyzing the results
along with prior data for $\alpha =1$, it is observed that the valley
induced by the particle-hole symmetry and the inversion symmetries in
alignment stacks, as well as the particle-hole symmetries in cyclic stacks,
lead to the disappearance of the Berry curvature associated with the flat
band. In contrast, the flat band is crucial in determining the orbital
magnetic moment in the alignment stacks. However, in the cyclic stacks, the
Berry curvature associated with the corrugated band becomes dominant at $k=0$%
. This corrugated band significantly enhances the value of the orbital
magnetic moment, which remains positive at $k=0$. The $\alpha -T_{3}$
bilayer lattice may enable the development of highly sensitive quantum
sensors that can detect subtle polarized radiation or magnetic fields by
altering material properties in response to these fields. This lattice also
has applications in optoelectronics, especially in secure communication
systems that encode information via light polarization, and in
valley-caloritronics. It presents new opportunities for quantum computers
and fast, efficient electronic devices by controlling electron properties at
the energy valley level, facilitating innovative information storage and
processing methods.

\section*{Acknowledgment}

The authors would like to acknowledge the "Acad\'{e}mie Hassan II des
Sciences et Techniques"-Morocco for its financial support. The authors also
thank the LPHE-MS, Faculty of Sciences, Mohammed V University in Rabat,
Morocco for the technical support through facilities.

\end{document}